# HUMAN-AI COORDINATION ZONES

## A Framework for Designing Human-in-the-Loop Experiences with Agentic AI

James Pierce

Amazon Web Services, jamsjp@amazon.com

Vaiva Kalnikaitė

Amazon Web Services, vaivakn@amazon.com

Siddharth Gupta

Amazon Web Services, guptsid@amazon.com

Brian Granger

Amazon Web Services, brgrange@amazon.com

As generative and agentic AI becomes embedded in everyday products, practitioners face a persistent challenge: how to design human-AI coordination—the ongoing mutual adjustment between users and AI systems as mediate through interfaces—that supports usability, trust, and safety. Existing resources offer high-level principles ("be transparent," "maintain user control") or low-level UI patterns, but there is a lack of mid-level design knowledge bridging the two. Through landscape and artifact analysis of 60 commercial AI applications, we introduce a framework defining human-AI coordination as the interplay of three dimensions: salience (how prominently AI is presented), involvement (what users can do to engage AI), and activity (what AI actually does). We contribute mid-level tools including coordination zones (done-for-me, done-under-me, done-with-me, done-without-me), an input taxonomy (prompted, sparked, inferred, layered), coordination curves for mapping user journeys, and design patterns demonstrating the generative capacity of the framework. The framework can be applied generatively to design experiences, analytically to evaluate existing ones, and communicatively to articulate ideas across stakeholders.

## 1 INTRODUCTION

Generative AI is increasingly embedded in products and experiences—from customer service to creative tools to autonomous agents capable of perceiving, reasoning, and acting independently over time [2][46][77][129]. Yet the design standards, user expectations, and best practices needed to guide these systems are still taking shape [134]. Foundational questions remain unresolved: How should humans and AI systems coordination? What levels of autonomy, transparency, and user involvement are appropriate for various contexts? How do we design for AI that may act independently, unpredictably, and invisible? These are not merely technical problems. They are design challenges that demand new conceptual tools.

To navigate these challenges, practitioners currently have access to two types of resources. High-level principles—such as "be transparent," [45][31][4] "design for appropriate trust," [128] and "keep users in control" [110]—articulate desired outcomes. Low-level UI patterns—like chat interfaces, confidence indicators, and prompt fields—provide clear and prescriptive implementation guidance.

What is missing is mid-level design knowledge [43][64]: actionable concepts that bridge *what-should* principles with *how-to* implementation. This gap is well-documented. Yildirim et al found that "while guidelines are useful for problem solving, practitioners desire better support for problem framing, ideation, and the selection of the right human-AI problem" [134]. Similarly, Naik et al concluded that practitioners struggled to implement "transparency, trust, and alignment" and suffered from "lack of [AI/agentic] standards" [83].

In our applied design work, we and our colleagues have observed that traditional methods like user stories, user flows, and existing design patterns and components often break down when designing for non-deterministic AI. This finding is consistent with recent reports from practitioners navigating similar challenges [134]. Without mid-level reference points, designers struggle to craft this new, generative material.

We use the term *coordination* rather than *interaction* to foreground a distinctive quality of agentic and generative AI: unlike conventional interactive systems where user actions produce predictable system reactions, AI-enabled experiences involve ongoing mutual adjustment between human and machine, where either party may initiate, adapt, or act independently. The concept of coordination—emphasizing mutual adjustment, shared awareness, and the management of interdependencies—has a long history in CSCW [74] and human factors [17][50][54], and we extend it here for UX design practice.

For example, when an AI scheduling agent autonomously books a meeting based on inferred preferences, the user is not simply "interacting" in any conventional sense. Yet the experience still requires coordination: the system must decide what to surface, the user must decide whether to intervene, and the interface must mediate between them.

More specifically, we define **human-AI coordination** as the ongoing relationship between a user and an AI system as mediated through an interface. We characterize this relationship through the interplay of three dimensions:

- **salience** (how prominently AI presents itself to and is perceived by users)
- **involvement** (what actions are available to users and how actively they engage)
- **activity** (what the AI actually does, regardless of user perception).

This definition foregrounds the interface layer—the space that many HCI and UX practitioners directly shape—rather than the underlying technical architecture. Within this deliberate scoping, we do not attempt to capture all aspects of human-AI relationships. Rather, we focus on the experiential and interaction dimensions most relevant to user experience design decisions. Throughout this paper, we use this definition to analyze 60 commercial AI applications and develop a practical vocabulary for describing and designing coordination across a spectrum—ranging from close collaboration to quiet background assistance to full automation.

Through landscape and artifact analysis of 60 commercial AI application and validated through applied design work at Amazon Web Services (AWS), we developed a set of mid-level concepts that are abstract enough to apply across contexts yet specific enough to guide implementation. Our contribution consists of a toolkit of operational concepts, including:

- **Coordination zones** (done-for-me, done-under-me, done-with-me, and done-without-me) that provide a flexible vocabulary for calibrating human-AI coordination.
- **Input types:** A taxonomy of conventional, prompted, sparked, inferred, and layered user input for AI.
- **Coordination curves:** Visual tools for mapping how coordination varies across user journeys.
- **Design patterns** (responsive salience, workplan gating, attribution markers, progressive autonomy) that demonstrate the framework's generative, analytic, and communicative capacities.

The framework can be applied generatively (to design new experiences and concepts), analytically (to evaluate and understand existing ones), and communicatively (to articulate ideas across stakeholders and communities). We conclude with a discussion of applications, limitations, and future directions.

**Human-AI Coordination**
What dimensions interact within coordinated experiences?
User Involvement | Salience | AI Activity

**Saliency Initiation**
How does AI become salient to users?
Sought salience (user pulls) | Imposed salience (AI pushes)

**Initiation Involvement**
How do input prompts vary user involvement?
Do-this | Do-that | Do-it | Already-done

**AI Input Types**
How do users provide input?
Conventional input | Prompted input | Sparked input | Inferred Input | Layered Input

**Coordination Zones**
What is the range of coordinated human-AI experiences?

**Done-with-me**
Mutually collaborative
Human is very in-the-loop

**Done-for-me**
Heavily automated
Human is barely in-the-loop

**Done-under-me**
Quietly assisted
Human is unawarely in-the-loop

**Done-without-me**
Covertly conducted
Human is out-of-the-loop

**Coordinated Work and Entities**
Who does what and when during a coordinated human-AI experience?

Phases
Initiate | Monitor | Update | Complete | Extend | Settings
Entities: User, UI, AI, Systems

**Coordination Curves and Typologies**
How does coordination vary with time?

Figure 1: Key components of our mid-level AI design knowledge toolkit. Basic concepts are the top, and more applicable design patterns and techniques towards the bottom.

## 2 MOTIVATION AND BACKROUND

### 2.1 AI as a Design Material

Researchers increasingly advocate for understanding AI and machine learning as a "design material" [7][10][20][89][132][133][135]. In this view, AI is something designers shape and work with, rather than simply implement. This perspective emphasizes that AI exhibits unique qualities such as non-deterministic output and uncertainty [7] which make it hard to work with [133]. Inspired by successful commercial AI systems like ChatGPT and Gemini, researchers advance novel demonstrations ranging from functional proof of concepts [16][65][68][69] to inspirational artifacts that showcase the aesthetic, participatory, and interactive potentials of AI [8][9][10][33][73][80].

Collectively, this work argues that designers need conceptual tools suited to AI's distinctive properties and cannot rely on straightforward applications of concepts from conventional software design. Prior work has attempted to articulate such properties. Dove et al [20] documented how UX practitioners struggle with AI's unpredictability. Yang et al [132] identified uncertainty, complexity, and evolving capabilities as core challenges. Benjamin et al highlight uncertainty as a quality with both negative and positive attributes [7].

This project contributes to this effort by focusing on the basic experiential qualities that designers can manipulate to channel, manage, and orchestrate such qualities. The dimensions of salience, involvement, and activity provide a vocabulary for defining and designing AI experiences that balances conceptual generality with practical applicability.

### 2.2 Guidelines for Human-AI Interaction

Researchers and companies have developed guidelines for creating effective, usable, and safe AI systems. A seminal 2019 paper by Amershi et al outlined 18 guidelines for human-AI interaction, such as "make clear what the system can do," "remember recent interactions," and "notify users about changes"[4]. Building on this work, Weisz et al consolidated six AI design principles that "address unique characteristics of generative AI: design responsibly, design for mental models, design for appropriate trust and reliance, design for generative variability, design for co-creation, and design for imperfection [128]. Major companies like Google [31], Microsoft [78], IBM [45], Meta [76], Amazon [1], Apple [3], and OpenAI [93] have published similar high-level AI principles, often focusing on ethics and responsibility. These projects extend an HCI tradition of guidelines and principles pioneered in the earlier days of interactive computing. Examples include Jakob Nielsen's 10 usability heuristics [85] and Ben Shneiderman's 8 Golden Rules of Interface Design [105].

Our work shares the goal of helping practitioners—designers, developers, and decision-makers—build useful, trustworthy, and safe AI products and services. However, we have found that high-level prescriptive guidance alone can be difficult to apply in practice, especially when it focuses on what should be achieved rather than techniques or elements for achieving. Guidelines like "design for appropriate trust" [128] are valuable, but they leave open the question of how to operationalize them in specific contexts and what and where practitioners can intervene.

To address this gap, we sought to develop what Höök and Löwgren [43] call "intermediate-level design knowledge"—practically applicable knowledge that is neither overly abstract nor overly prescriptive. In this tradition, our framework does not deliver low-level implementation patterns or plug-and-play UI elements. Instead, it provides scaffolding for generating and organizing such patterns. Our framework functions as a vocabulary and structure for articulating coordination decisions and techniques. We return to this generative capacity and its application to future work in the Discussion.

### 2.3 Coordination and Related Concepts

HCI, human-AI interaction, and human factors have developed numerous concepts for understanding how humans and AI systems can work together. Our framework draws on and extends several of the traditions.

Levels of automation (LOA) taxonomies describe the division of labor between humans and machines along a spectrum from fully manual to fully autonomous [95][109]. LOA frameworks have been influential in human factors engineering, but they primarily characterize what they system does—the degree of automation—rather than how that automation ius presented to and experienced by users through an interface. Two systems with the same levels of automation may offer very different user experiences depending on how visible (salient) and controllable (involvement) the AI's activity is.

*Human-in-the-loop (HITL)* approaches emphasize human involvement in AI oversight and decision-making [18][22][130]. HITL has become a widely used shorthand, but it is often treated as binary: the human is either in the loop or not. In practice, users may be involved in many different ways—approving a plan, monitoring progress, adjusting parameter, or simply being aware that AI is acting—and the gradations matter profoundly for user experience and interface design.

*Explainable AI (XAI)* and *transparency* research focuses on making AI reasoning and operation understandable to users [38][61]. These concepts address what information is made available, but the concepts alone do not fully capture whether users actually attend to, perceive, or act on that information in context—a gap our concept salience and involvement is designed to address.

*Proactivity and mixed-initiative interaction* examine how AI systems can take initiative while preserving human control [44][136]. These concepts describe system capabilities but may, in some treatments, leave open the question of how proactive behavior is experienced and mediated through the interface.

*Coordination in human-agent systems* draws on CSCW and human factors traditions emphasizing *observability* and *directability* as key properties of effective human-agent teaming [50][54]. Our framework builds directly on this tradition, translating these concepts into dimensions that UX designers can operationalize.

Each of these traditions captures important aspects of human-AI interaction. However, they tend to address different facets of the design challenge—automation level, human oversight, system transparency, initiative—but without a shared vocabulary that connects that the level of the user interface design and user experiences. Our framework aims to provide such a vocabulary, organized around three dimensions (salience, involvement, and activity) that we found useful for relating these concepts to one another in design practice. We return to these relationships to prior work in detail in Section 6, after presenting the framework.

## 3 METHODS

Our method combined landscape and artifact analysis of a curated corpus of 60 commercially available AI software applications. We sought to create practical conceptual tools to assist designers and decision-makers in articulating and implementing appropriate forms of human-AI coordination across different contexts. While many disparate collections of AI UX patterns and principles exist, we found a lack of taxonomies and principles to organize this guidance, help practitioners navigate it, and function as scaffolding for generating new design patterns, principles, and terminology. We further identified a need to articulate a range of human-in-the-loop engagements, rather than treating this as a binary, all-or-nothing design consideration.

### 3.1 Framework Scope and Requirements

In developing the framework, we were guided by three goals: (1) Descriptive clarity: The framework should help practitioners and researchers articulate different types and degrees of human-AI coordination, and be widely applicable and adaptable to different contexts. (2) Practical applicability: The framework should support design, development, evaluation, and decision-making—particularly in cases of non-deterministic and autonomous agentic AI. (3) Generative extensibility: The framework should serve as scaffolding for creating new AI UX design knowledge, including patterns, principles, guidelines, terminology, and design exemplars. We also clarified an important scope boundary: (4) higher-level guidance, not low-level implementation. Our framework does not prescribe specific design patterns or UI elements. Instead, it is designed to help explain, organize, and generate such knowledge—as we discuss further in the Discussion.

### 3.2 Process Overview

Our process consisted of four phases. First, we conducted a landscape analysis of existing AI UX resources. Second, we collected a corpus of 60 AI applications. Third, we analyzed these applications using exploratory and focused techniques. Fourth, we synthesized our findings into taxonomies and definitions. Throughout, we validated our emerging framework with internal stakeholders at AWS and by applying our ideas to real-world design challenges. Prior work has employed similar approaches of combining artifact analysis, expert review, and internal validation (e.g., [71][72][103]).

Our framework was iteratively validated through three forms of internal expertise and external feedback. **(1) Core research team validation.** Our core research team consists of three senior UX designers, four software developers (three junior, one senior), a principal technologist, two senior leaders (UX and software development), and a visiting professor specializing in HCI, design, and digital trust. We drew on this combined expertise throughout the process.

**(2) External consultation.** We closely consulted 10 additional experts across UX practice, software development, product management, and industry research. Key engagements included:

- A 90-minute formal feedback session with 6 UX designers during a weekly design crit at AWS.
- Reviews with two senior scientists—one in responsible AI, another in open-source software development.
- A presentation to approximately 100 in-person attendees (plus many additional remote participants) at an internal conference, which generated dozens of follow-up requests to share our work and discuss it.

**(3) Applied validation.** We applied the framework to three projects within our organization: (1) R&D prototypes exploring novel AI UX patterns for user trust, control, and safety; (2) AI UX design for a specific enterprise product used globally by thousands of users; and (3) prescriptive guidance documents used across a large technology company. Details of this work are not included in this reporting.

### 3.3 Landscape Analysis: Aggregating New AI UX Terminology, Patterns, and Principles

In phase one, we built an internal AI UX design knowledge base for AWS. This initial work focused on exploring and curating AI design resources to assist designers and decision-makers across our company, though we had not yet refined our focus on human-AI/agent coordination challenges. We compiled terminology, principles, guidelines, and design patterns from multiple sources, including online pattern libraries (e.g.,[107]), articles by and for UX practitioners (e.g.,[51][108]), and research publications (e.g.,[52][83][96]). (See Appendix A.1.3 and A1.1.4 for sources and samples).

A key finding was a lack of novel mid-level design patterns specific to generative AI. Most guidebooks and pattern libraries contained high-level principles or low-level component patterns (e.g., “filters,” “file attachments for AI prompts,” “nudges for AI features”) that are simple extensions of existing UI patterns.

Phase one culminated with sharing our AI UX Knowledge Base with seven internal stakeholders (one AI director, one principal technologist, one UX manager, and four UX designers) who provided feedback and helped us validate and refine our direction and then refocus our research goals. For example, a team at our organization validated the need for actionable design guidance that goes deeper than high-level principles. Another team described how standard methods they relied on for design and development, like user stories and user flows, did not work as well when designing highly unpredictable AI/agentic systems.

From press releases and news stories, we found that many at our organization and outside were focused on a vision of fully automated agents. Yet we saw an inevitable need for human involvement, even if that involvement was minimal or rare. This insight revealed a need for better tools to design appropriate levels of human oversight, particularly for agentic AI systems. As a result, we narrowed our focus to human-in-the-loop interaction with an emphasis on agentic AI.

The landscape analysis yielded three findings that shaped the rest of our research. First, existing AI design pattern libraries overwhelmingly focused on chat-based interfaces and low-level UI components (e.g., prompt fields, confidence indicators, loading states), with little guidance for the broader range of new and emerging coordination patterns we observed in agentic systems. Second, the terminology used across industry guidelines, research publications, and practitioner resources was fragmented. Concepts like transparency, explainability, human-in-the-loop, and levels of automation were discussed in isolation, but without a shared vocabulary connecting them at the level of user interface and experience design (UX/UI). Third, the practitioners we consulted confirmed that this fragmentation was not merely semantic or academic—it created real difficulties in formulating design solutions and communicating design intent across teams. These findings motivated our decision to develop a coordination-focused framework grounded in the analysis of real applications—and with an eye toward future evolutions of the field.

### 3.4 Collecting AI Examples

In phase two, we collected a corpus of 60 applications employing AI. (See Appendix A.1.1 for the full list). Our goal was to gather a diverse set of AI UX examples for analysis, spanning different use cases, interaction models, and degrees of AI prominence. Our corpus included popular tools like OpenAI's Deep Research and Operator, and AI features in Adobe Photoshop, Canva, and Figma, along with significantly less popular and effective tools.

We applied the following inclusion criteria:

- **Access.** We limited our corpus to consumer applications we could easily access, and excluded most enterprise software, and all expensive niche applications and experimental prototypes.
- **AI type.** We focused on generative and agentic AI but selectively expanded to include non-LLM implementations of AI autonomy and proactivity. We found that current consumer applications cluster around certain patterns (e.g., chat interfaces, collaborative experiences), and we sought greater variation to uncover emerging design directions for AI UX. This necessitated expanding beyond openly LLM-based AI applications.
- **Variance**. We sought diversity across multiple dimensions: use case, AI prominence (is AI integral to the user experience or supplemental?), level of AI autonomy and proactivity, potential risk to users, and UI implementation (we sought both common and novel features).
- **Interaction models beyond chat**. We actively searched for applications that did not conform to the dominant chat interface model employed by general-purpose AI tools like ChatGPT, Claude, and Gemini.

### 3.5 Analyzing AI UI Techniques and Experiences

In phase three, we conducted a two-part analysis of our corpus: an initial exploratory stage followed by a focused analysis using more structured techniques.

**Part 1: Exploratory analysis.** We began with open-ended exploration to immerse ourselves in examples and understand real-world applications of new and emerging (1) AI design patterns (e.g., "workplans," "show your work"), (2) UI elements (e.g., pause/improve prompt buttons), and (3) adherence to AI UX principles (e.g., "explain the results").

As an initial immersion activity, we conducted a 60-minute workshop with 8 practicing UX designers—each with at least five years of professional experience. Each participant reviewed up to three applications from our curated corpus. Participants were asked to capture noteworthy features, patterns, elements, and moments—both positive and negative— and document their observations with annotated screenshots and GIFs. This workshop yielded a scannable initial snapshot of AI experiences across the corpus. The first author then reviewed every application in the corpus, using these initial snapshots to help determine focus and priority.

Our exploratory analysis was guided by several considerations that emerged through deep engagement with examples and reflective annotation:

- **Outliers.** Given our goal of understanding the range of AI UX approaches, we noted qualities and techniques not captured in the existing guidance we curated in Phase 1. Examples include robust handling of user interruptions, generative conversion of text to interactive menus within chat dialogues, and varying approaches to AI attribution through micro-copy, placement, and iconography (e.g., "This review was generated by AI.").
- **Representation.** We attended to how AI was represented in the UI: contrasted versus muted, integrated versus separate, defining versus accented.
- **Qualities.** We began synthesizing quality dimensions useful for differentiating AI experiences, including prominence, proactivity, transparency, predictability, task complexity, and review effort. These eventually consolidated into our three-part coordination framework: salience, involvement, and AI activity.

From this exploratory work, we began synthesizing core framework elements: a "layers of AI user journey" diagram (phases + actors), how coordination varies over time across user journeys, and the simplified coordination dimensions.

**Part 2: Focused Analysis.** In the second portion, we selectively analyzed specific products to examine how they performed across coordination dimensions and over time. (See Appendix A.1.2 for details). We employed three techniques:

- **Annotated screenshots.** We captured static screenshots and GIFs, and then annotated them with existing AI UX terminology (patterns, UI elements, principles, interaction models). We created new terms when none existed. See Figure 2 for a sample.
- **Coordination curve Mapping.** Using our layers diagram (Section 4.4), we developed a tool of plotting user journeys called coordination curves (Section 4.5). We plotted selected examples according to salience and involvement across six phases (initiating, monitoring, updating, completing, extending, and settings) and noted which actors were active at each point in the journey (user, UI, agent/AI, other systems).
- **Combined analysis.** We then merged the two methods by extracting UI screenshots for user journeys and placing them below our coordination curve plots, where we then annotated them with identified features and patterns.

### 3.6 Synthesizing Taxonomies and Definitions

In phase four, we synthesized our findings into a set of taxonomies and definitions. To direct our analysis toward synthesized outcomes, we used a running document for capturing and iteratively refining emergent themes and concepts. Our primary analytical approach was artifact analysis: systematic examination of design artifacts—screenshots, interaction flows, and UI elements—to identify recurring patterns, qualities, and relationships [11][28][41][90][99]. Concretely, this involved iteratively annotating screenshots with emergent codes describing UI elements, patterns, goals, and other mechanisms. For example, we iteratively annotated UI helper text and annotations announcing AI limitations, guardrails, AI-generated content leading us to the general label of "AI annotation layers" (See Figure 2 for examples). We then clustered codes into emergent themes, and refined those themes against the full corpus.

This process drew on thematic analysis techniques [32][79] adapted for artifacts rather than interview data. We also iteratively applied our emerging framework to real-world design projects at [large technology company], which served as a practical check on the utility of our constructs, though we do not present details of those applications in this paper. Our analytical lens was shaped by HCI research on design qualities and materials (e.g., [7][26][71][118][123]), and studies of design and UX practice (e.g., [35][36][37][52][56][112][117][133][139]).

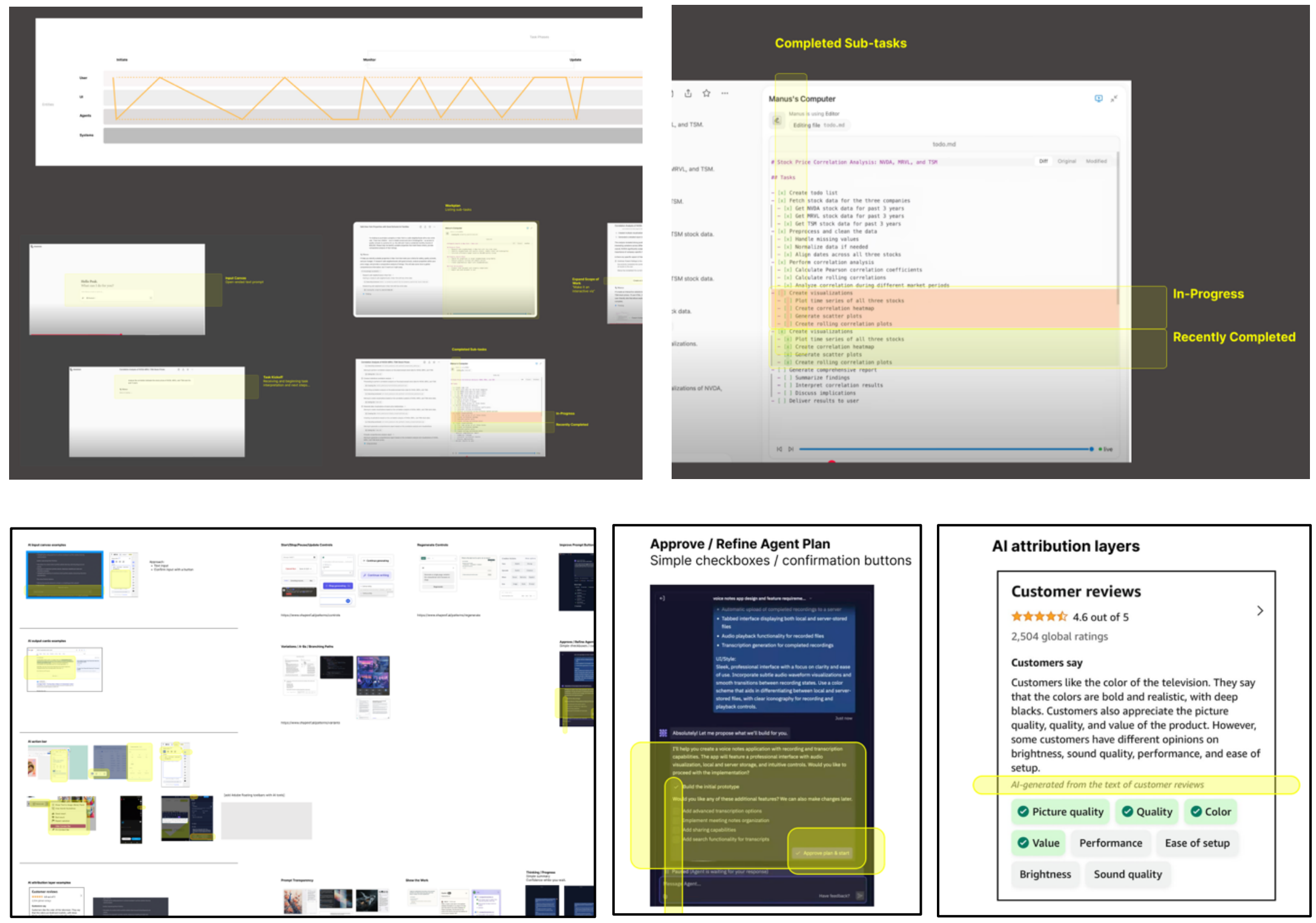

Figure 2: Examples of our analysis of existing AI features and patterns. (Top) An example of how we analyzed existing AI applications. We mapped the coordination curves for typical user journeys by arraying the four entity layers (user, UI, agent/AI, other systems) and the six main phases (initiating, monitoring, updating, completing, extending, and settings). We then added UI screens to the coordination curve plot, and annotated these with labels highlight features and patterns we identified. This sample is of Manus, a general-purpose agentic AI tool. (Bottom) Examples of our analysis of individual UI screens for AI applications to extract recurring design patterns and new AI UI components. Samples are shown for patterns of workplan approval and refinement (with Replit AI software development tool) and AI attribution layers (for Amazon e-commerce).

## 4 THE FRAMEWORK: CONCEPTUAL BUILDING BLOCKS FOR HUMAN-AI COORDINATION

One of the most critical decisions in AI product design is determining how visible and interactive AI capabilities should be. Should users actively direct it, let it act autonomously, or somewhere in between? And how should this change based on the task, the user, or over time? This section introduces a vocabulary for defining human-AI interaction. These concepts help describe and design the right types and degrees of "human-in-the-loop" for varying contexts.

We begin by defining human-AI coordination as the interplay of three dimensions: salience, involvement and activity (4.1). We then introduce four coordination zones that characterize common human-AI experiences (4.2), followed by a taxonomy of UX input types (4.3), a model for articulating how coordination varies by time and scale (4.4), and coordination curves for visualizing user journey and recurring temporal coordination patterns (4.5-4.6).

### 4.1 Defining Human-AI Coordination

With conventional interactive systems, user actions lead to predictable system reactions. Agentic AI differs in that inputs and outputs are less tightly coupled, agents may act proactively without direct user initiation, and interaction often resembles an ongoing correspondence rather than discrete action-reaction cycles. For these reasons, we find it useful to shift from the concept of *interaction* to *coordination*—a term that better captures the mutual, evolving relationships between humans and AI systems. Coordination occurs within the envelope of interactivity and does not replace it.

We define *human-AI coordination* as the interplay of three dimensions of human experience and technical activity:

- **Human involvement:** The degree and forms of effort a user invests, or is demanded of them, in working with an AI system via the user interface. Involvement is constrained and enabled by user interface options and affordances, and thus directly connects to users' perception of AI capabilities via the user interface.
- **Salience:** The degree and forms with which a user perceives AI via the interface. Salience describes how prominently the AI is presented to users. More visible, vocal, proactive, or transparent AI will have comparatively higher salience. Salience is an important dimension of many recurring qualities of great importance to AI user experiences, including transparency, explainability, visibility, and proactivity [4][96][128]. Pu et al, for example, recommend adjusting the "salience" (defined as "prominence") of "proactive agent actions" based on task significance [96]. This is a type of design insight and hypotheses that our framework seeks to operationalize.
- **AI activity:** What the AI actually does, regardless of how much is exposed to users. In UX-oriented practice, we find that the focus is often on involvement and salience, as activity is typically experienced by users only through its salient manifestations. However, total activity remains conceptually important and a critical dimension in user trust: users often form mental models of underlying AI behavior that may or may not reflect the actual underlying system activity.

These dimensions combine to define varying degrees of coordination. When both involvement and salience are high, coordination is high, and represents a collaborative, "in-the-loop" experience. When both are low, coordination is minimal, and represents an automated, "out-of-the-loop" experience. This **ratio** of involvement to salience provides the foundation for our coordination zones, described next. Each dimension can be decomposed into more granular concepts. For example, involvement includes user actions such a directing, requesting, overriding, pausing, approving, or sustained monitoring. Salience may include user perceptions such as noticing alerts, interpreting explanations, tracking progress, or comprehending AI reasoning.

**Why these three dimensions?** We suggest that important qualities—such as predictability, trustworthiness, explainability, and proactivity/agency—are shaped in part by the interplay of salience, involvement, and activity: three key levers that designers can manipulate. For example, a system may become more predictable when users can observe its behavior (salience) and learn how their actions (involvement) relate to outcomes. This is a working hypothesis that we find productive for framing design concepts and decisions, though it warrants further empirical investigation. We do not claim these three dimensions fully determine such qualities, only that they provide useful conceptual levers for designers to consider.

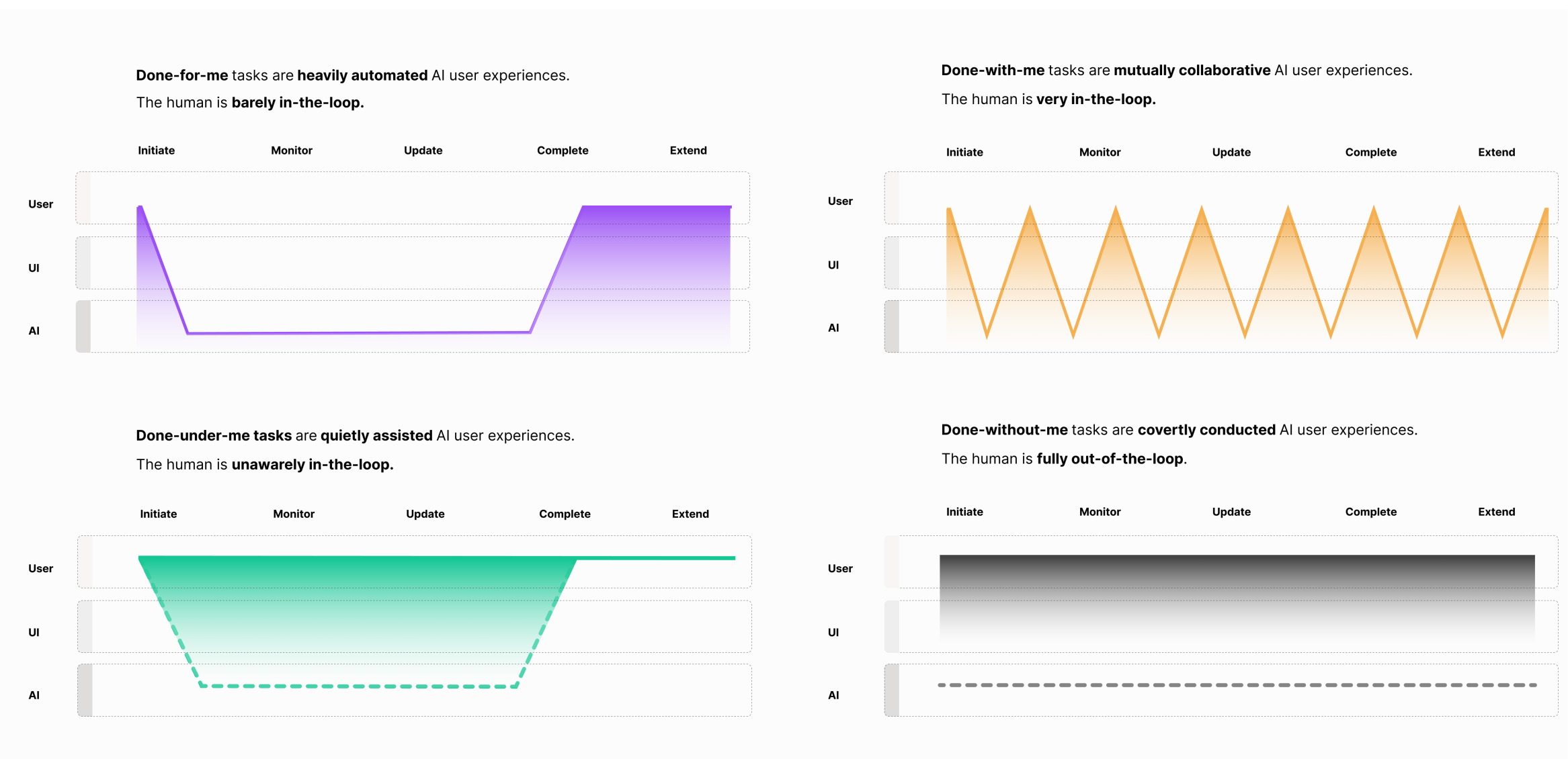

Figure 3. Typical coordination curves for the four basic coordination zones: Done-for-me, -by-me, -with-me, and -without-me.

Second, these dimensions have grounding in prior work. Coactive theories of automation argue that *observability and directability* are essential for human-AI coordination [50]. Salience and involvement adapt these ideas for HCI and UX practice. Salience and involvement also map to *perception and action* [124], which are fundamental dimensions of human experience with particular importance for interface design. We add AI activity to address potential mismatches between what users perceive AI is doing and what it actually does.

Third, our 3-part framework is simple, memorable, and widely applicable. The coordination zones we describe next are explained by plotting salience against involvement, with activity as a third dimension capturing underlying system behavior. Indeed, the dimensions emerged empirically as we sought to understand what distinguishes collaborative from automated experiences, and where AI is prominently surfaced versus quietly backgrounded.

### 4.2 Coordination Zones: Done-For, Done-Under, Done-With, and Done-Without-Me

Many discussions of agentic AI reflect binary thinking: the human either is, or is not, "in the loop." A key insight of our framework is that human-AI coordination exists on a spectrum, with many gradations between a fully involved user working with a highly salient AI and an uninvolved user unaware of background AI activity. This concepts extends thinking from levels of automation frameworks in human factors engineering [91][109][122], but focuses on user experience and interface mediations, rather than the resulting automation of activity.

Coordination typically varies throughout a user experience. A user may be highly involved during task initiation but uninvolved during AI execution. Conversely, an agent may proactively complete work that a user later reviews and extends. Even highly autonomous "background" agents require occasional human coordination during setup, approval, or auditing.

Our framework articulates four basic zones of human-AI coordination (Figure 3). These range from highly collaborative *done-with-me* experiences to *done-for-me*, *done-under-me*, and *done-without-me* interactions where involvement or salience diminish. This is not a rigid taxonomy but a flexible vocabulary for calibrating the ratio of human involvement to AI salience for specific design contexts.

#### *4.2.1 Done-For-Me (Heavily Automated)*

*Done-for-me work* is completed by AI with minimal user input. AI salience is typically moderate to low and is concentrated at initiation and/or completion phases. During execution, the AI may be entirely backgrounded. We can think of the user as *barely in-the-loop*.

Apple Watch's workout auto-detection illustrates this zone. The system infers from sensor data that the user is exercising, surfaces a brief notification ("it looks like you're working out"), and requires only a single tap to confirm or dismiss. The user's involvement is concentrated in the momentary approval or rejection. The AI proactively handles detection, tacking, and logging autonomously.

Done-for-me experiences range from user-initiated requests to fully proactive AI delivery. The degree of user involvement at initiation can vary significantly, spanning from detailed prompting to simple approval of an AI-prepared plan to pre-approved autonomous action. We describe the initiation spectrum later in section 4.3.2.

#### *4.2.2 Done-Under-Me (Quietly Assisted)*

*Done-under-me* work is performed by the user with quiet AI assistance operating in the background. AI salience is low or absent, and users may not consciously register AI activity. We can think of the user as *implicitly in-the-loop.*

Netflix's recommendation interface exemplifies this zone. AI curates content based on viewing history and inferred preferences, yet the interface presents recommendations as straightforward browsing categories. Users experience their journey as freely choosing options, while the AI's role in shaping what is available remains largely invisible.

Done-under-me experiences often share the following characteristics: the AI works without announcing itself (**done quietly**), acts rapidly enough that users barely register involvement (**done quickly**), and delivers outcomes users can immediately understand (**done clearly**).

#### *4.2.3 Done-With-Me (Mutually Collaborative)*

Done-with-me work involves close collaboration between user and AI across multiple phases—initiation, monitoring, updating, completion, and extension. AI salience is moderate to high, and human involvement is significant. We can think of the user as *very in-the-loop*. Done-with-me experiences represent the most balanced form of coordination, where neither party dominates.

Manus AI's agentic workflow exemplifies this zone. The system presents a visual workplan that updates dynamically, shows its activity in real time (e.g., browsing, writing code), and supports mid-execution interruption and modification. The user's involvement peaks at initiation and at key checkpoints, with moderate engagement during execution. This illustrates an iterative back-and-forth that distinguishes canonical done-with-me experiences from the brief touchpoints of done-for-me interactions.

#### *4.2.4 Done-Without-Me (Covertly Conducted)*

*Done-without-me* work is conducted by AI without user awareness or involvement. The user is *completely out-of-the-loop*. While pure done-without-me experiences largely fall outside our research scope—as they do not directly involve

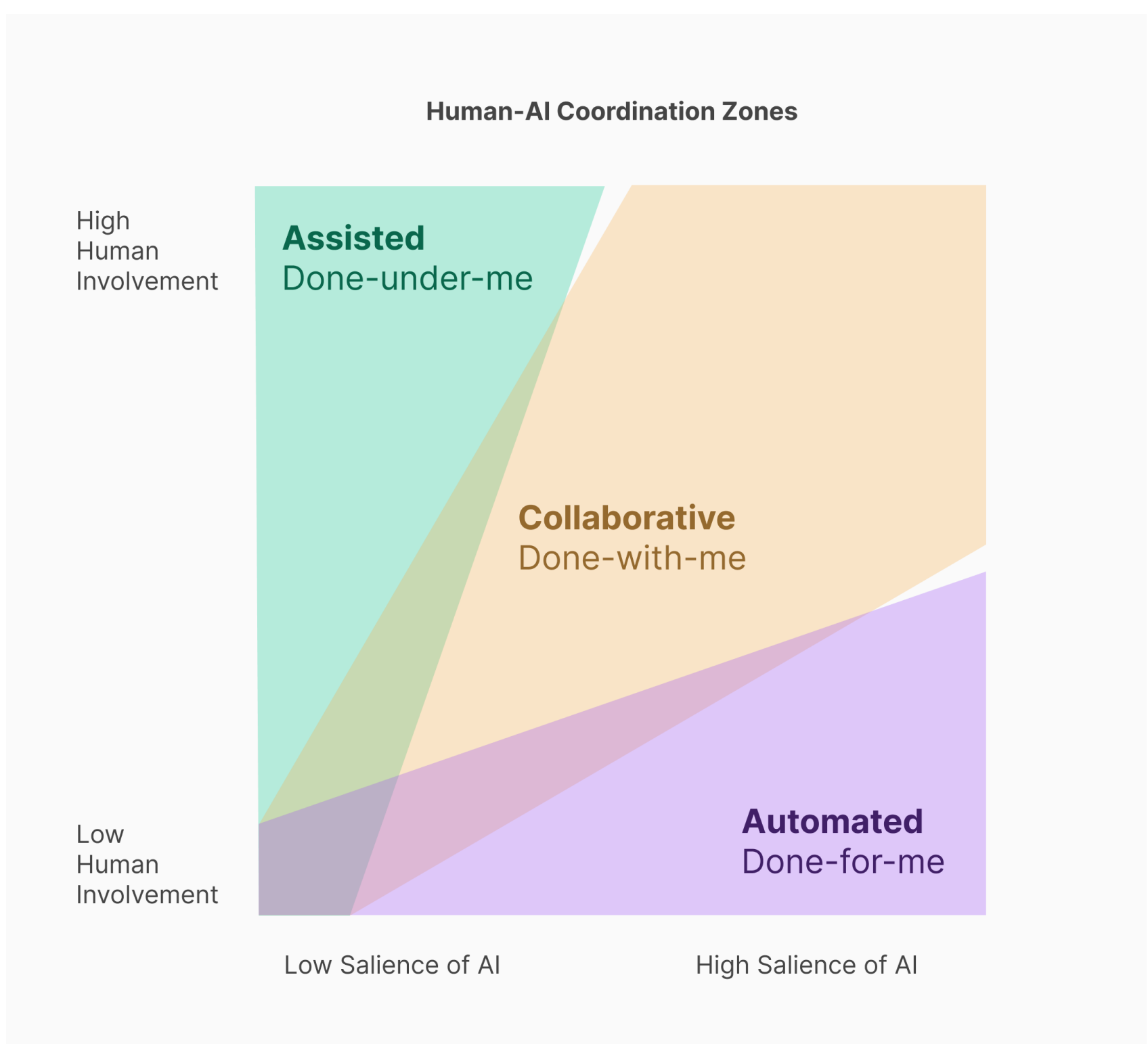

Figure 4: This diagrams plots done-by-me, done-for-me, and done-with-me experiences according to the levels of agent salience and human involvement. From this visualization, we can see that done-by-me (quietly assisted) experiences have low agent salience and variable human involvement, done-for-me (heavily automated) experiences have low human involvement and variable agent salience, and done-with-me (mutually collaborative) experiences are characterized by parity in the level of agent salience and human involvement but typically involve higher levels of salience and involvement.

end-users—they remain important to consider. Users often form mental models that include beliefs about hidden AI activity, whether accurate or not [19][30][83][96]. Perceptions of covert AI behavior—such as extracting sensitive data or performing unauthorized actions—can significantly affect trust, even when such activity is not occurring [37][39][140]. Designers should be mindful of how their systems might be perceived to users, not just how they technically function.

### 4.3 Input

Having defined coordination zones, we now turn to the input mechanisms through which humans can direct, steer, and control AI within a coordinated experience. How users provide input is a crucial component of each coordination zone.

*4.3.1 User Input Types*

AI systems enable new ways for users to provide input beyond conventional UI elements. These input modalities can be combined with traditional interfaces to create effective coordination experiences.

**Conventional Input.** Traditional UI elements—buttons, menus, text fields, links—follow a deterministic action-reaction pattern. Users act, and systems respond predictably. These elements remain foundational in AI systems but are increasingly augmented by AI-specific input modes.

**Prompted Input.** Prompting occurs when users deliberately interact with a designated *AI input canvas* and receive generative outputs such as chat responses or content cards. Prompts can be formed by a combination of:

- **Action prompts** ("do something"): Requests for tasks or outputs, such as answering questions, generating content, or executing commands.
- **Context prompts** ("know something"): Information the user provides—such as examples, preferences, or background—that do not necessarily request immediate action.
- **Guide prompts** ("follow something"): Constraints, rules, or boundaries the AI should respect, such as tone, format, or guardrails.

**Sparked Input.** Sparking occurs when user input generates an AI response that qualitatively breaks from expected or established output patterns, presenting novel forms of UI elements. Unlike the variability inherent in all generative AI, sparking involves an abrupt or unpredictable shift in the *type* of response that is qualitatively different, and it is typically represented by a new and distinctive UI element.

We distinguish two forms sparking. *Cold sparking* occurs when a conventional-appearing UI experience—a button click, hover, or scroll—triggers a generative AI response. For example, a search query might produce an AI-generated summary rather than a list of links. This pattern can enable done-under-me experiences where AI assistance is suddenly made salient, and abruptly transitions to a done-for-me or done-with-me experience.

*Warm sparking* occurs within a prompted AI experience when the response is of a noticeably different kind than typical or predictable output formats. For example, an AI chat that usually produces purely text responses may occasionally spark distinct UI elements and actions—such as an indicator that the system is searching the web, a note that input has been saved for reference, or a message that the AI is "thinking longer," with options to stop or resume standard processing. The shift is experienced not as a change in degree but a change in kind.

In both forms, sparking disrupts user expectations about what a given input will produce. This can create welcome and seamless assistance, or, if poorly calibrated, confusion and frustration.

**Inferred Input**. Inferring occurs when systems extract information from user activity or context to trigger AI activity, often with low or no salience. For example, an AI might infer user intent from browsing patterns or detect confusion from interaction timing. Unlike sparking or prompting, inferred input can operate invisibly, enabling quiet assistance (done-under-me) but also raising concerns about covert activity (done-without-me).

For example, Apple Watch extracts exercise signals from accelerometer and heard rate data without an explicit user prompt, while Netflix infers viewing preferences from watch history and browsing patterns. In both cases, the user's ordinary activity becomes key inputs that drive AI activity.

**Layered Input.** These modalities can, and typically are, combined. For example, inferred input might assess a user's apparent trust level, which then informs how much transparency (salience) the system provides in prompted interactions.

These input types form a spectrum of user awareness and deliberation—from conscious and consensual (prompting), to less explicit (sparking), to potentially imperceptible (inferred). Input modes tend to align with coordination zones. Done-with-me experiences rely heavily on prompting. Done-under-me experiences typically depend on inferred input. Sparking bridges zones: cold sparking can abruptly transition users from quiet assistance into salient done-for-me or done-with-me experiences. Layered input combining these modalities is increasingly common in contemporary AI applications, including the use of sparking to intelligently determine between outputting do-this, do-that, do-it, or already-doing UI options.

#### *4.3.2 Initiation Involvement*

Prompted interactions may range from high to low involvement initiation. We identify four main types of prompt interfaces representing a spectrum of involvement:

- **Do-this**: User specifies the task in detail via open text or speech. Example: Detailed verbal prompts.
- **Do-that**: User selects from AI-generated options. Example: AI generated menu selection.
- **Do-it**: User approves or rejects an AI-prepared plan. Example: Accept/reject plan buttons.
- **Already-done.** AI acts proactively with pre-approved permission. Example: Pre-approved background actions.

The *already-done* category raises particular issues. Users may experience it as a prior pre-approval they granted, or—in extreme cases—covert activity occurring without meaningful consent. This initiation spectrum thus surfaces critical UX qualities impacted by initiation type, including user intent, awareness, consent, and autonomy. See Appendix A0 (Table 1) for descriptions with examples.

#### *4.3.3 Salience Initiation: Sought versus Imposed*

Who initiates salience AI activity? There are two basic ways that salient AI activity can be initiated:

- **Sought Salience** (human pull). Users request or search for AI inputs, outputs, or capabilities. The agent is made available based on user input requests or motivated search.
- **Imposed Salience** (AI push). The AI proactively increases its presence through, for example, notifications, suggestions, animations, or blocking modals.

Designers can sometimes cleanly choose one pattern over the other, though many interactions blur the distinction.

### 4.4 Coordinated Work: Phases, Scales, and Entities

Human-AI coordination unfolds across time, involves multiple actors, and occurs at varying scales of work. Naming and visualizing these dimensions can help designers manage the complexity and non-determinism inherent in agentic systems. This portion of the framework attempts to capture the real-world complexity of coordinated work balanced with the need for operational simplicity.[1] We organize coordinated work with regard to phases and scales. See Figure 5 for a diagrammatic visualization.

#### *4.4.1 Phases*

Coordinated work moves through characteristic phases where human involvement and AI salience may vary:

[1] For example, many real-world use cases can be characterized as **exploratory workflows**. A great deal of the work that humans perform involves open-ended inquiry and experimentation. One attribute of exploratory work is that the user does not yet a have a clear and stable goal, envisioned outcome, or success criteria in mind. Interacting with agents/AI helps the user figure out what they're trying to achieve and what is achievable. A common example is researching a topic with a conversational agent, like GPT, Claude, or Gemini chat applications. In a typical workflow, the user's workstream is emergently composed of many successive done-for-me tasks (e.g., "what is X?" "is X better than Y?", "based on my preferences, do this" "explain more and include sources"). Each done-for-me task is initiated relatively quickly by the user and then quickly completed by the agent. In aggregate, the automated done-for-me tasks form a highly collaborative done-with-me workflow.

- **Initiation:** The beginning of a task or workflow. Users may prompt, select, approve, or simply observe AI-initiated work.
- **Execution:** The middle phase where AI performs work. This may include:
    - **Monitoring:** Users observe AI progress.
    - **Updating:** Users revise prompts, task parameters, or workplans mid-execution.
- **Completion:** AI delivers output as the culmination of a task.
- **Extension:** Follow-up activities such as sharing outputs, requesting revisions, or initiating related tasks.
- **Setting and Auditing:** Parameters that govern coordination, either *pre-initiation* (configuring preferences, rules, permissions) or *post-completion* (reviewing logs, analyzing workflow history).

A *composite workflow* consists of multiple sequential or parallel phases. For example, an AI may complete an initial task, surface results for user review, then continue into a second phase based on user feedback. A typical engagement with an AI chatbot, for example, will often consist of multiple prompt (initiation), output (completion), review (extension), followed by a new prompt (re-initiation). As this example reveals, it can be useful—even necessary—to consider coordinated work according to varying scales.

*4.4.2 Scales*

Coordination needs differ by scale, and framing design situations according to different scales yields different design responses. We distinguish five nested levels to consider:

- **Microinteractions:** Momentary UI events, such as feedback indicators, confirmations, or brief guidance.
- **Tasks:** Discrete work units with clear initiation, execution, and completion (when successful).
- **Workflows:** Multiple tasks in sequence or parallel directed toward related goals.
- **Workstreams:** Collections of related workflows within broader work.
- **Work practices:** The broader role or occupation of a user, such as “software development” or “shopping.”

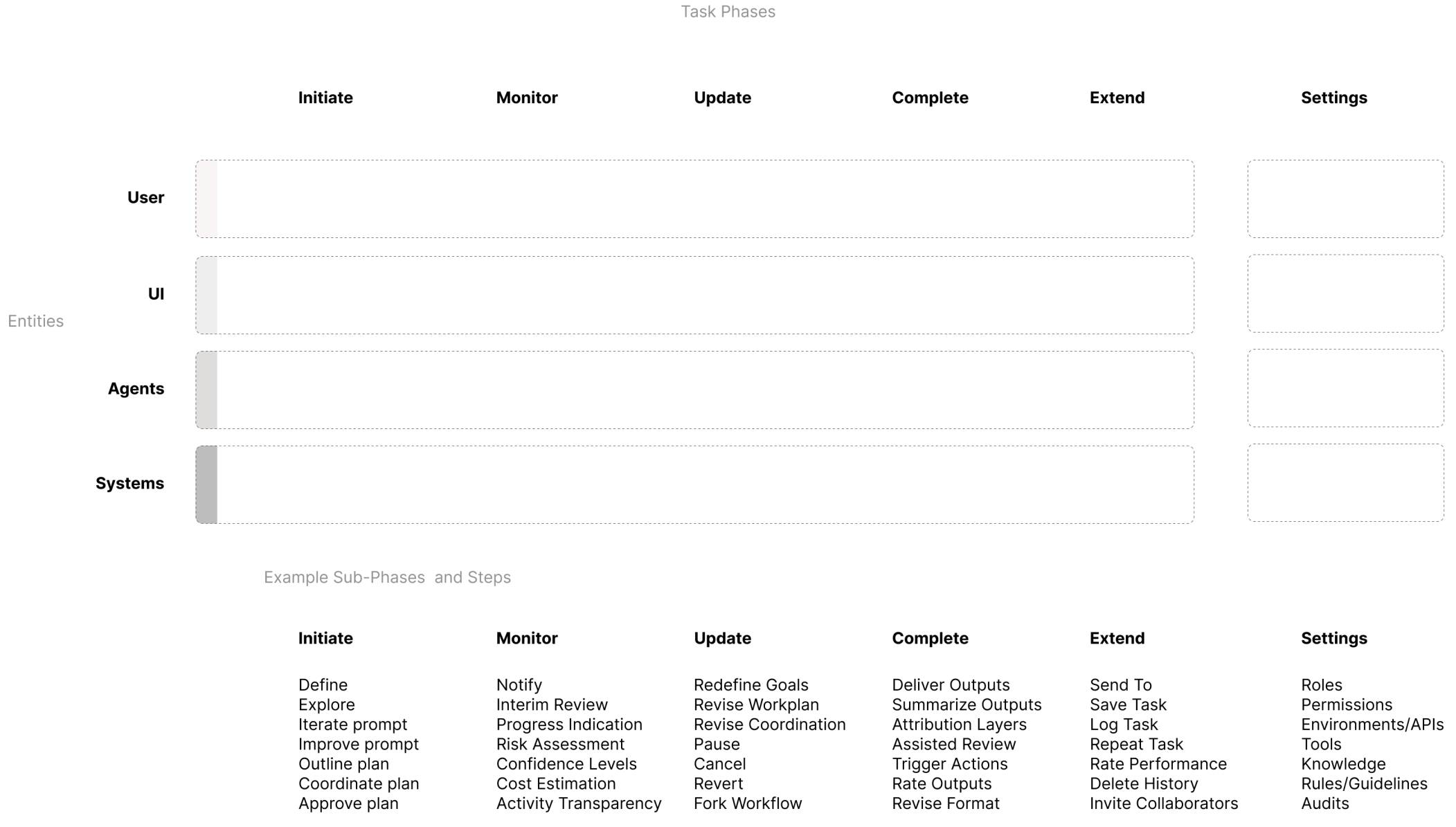

Figure 5**.** Human-AI work phases and layers (top) and examples of specific sub-phases and steps (below).

#### *4.4.3 Composite Work*

Many real-world workflows are composites of different coordination types. Scale is a crucial consideration in properly understanding these experiences. Consider an example of GPS navigation. Selecting a route may be done-with-me (user chooses among options). Turn-by-turn guidance is done-for-me (system automatically announces and directs, so the user can follow). Automatic re-routing is done-under-me (AI quickly, clearly, and quietly adjusts). Data analytics may be collected for insurance surveillance or vehicle diagnostics (an invisible and largely done-without-me activity).

Design choices can be applied to shift the coordination profile. If an AI provides transparent reasoning and supports mid-task interruption, an automated done-for-me task becomes more collaboratively done-with-me. Conversely, a done-for-me workflow may still require a done-with-me initiation phase to establish the work request and plan.

### 4.5 Coordination Curves and Typologies

*Human-AI coordination (HAC)* curves visualize how involvement and salience vary over time. They provide a visual summary of a user journey across coordination zones. Unlike conventional interactive systems, AI-enabled experiences may include significant periods where users are uninvolved, yet AI is conducting sophisticated work. These periods may require notification, approval, monitoring, or correction. These moments appear as “valleys” in a coordination curve.

For instance, mapping OpenAI’s Deep Research application reveals a characteristic curve shape: a high peak of involvement at initiation (where the user crafts a detailed prompt and the system asks clarifying questions), a sustained valley during the autonomous execution (where the user has limited ability to monitor, update, or otherwise intervene), and a second peak a completion (where the user reviews, evaluations, or extends the output.) By contrast, Manus AI’s curve shows multiple intermediate peaks corresponding to workplan checkpoints where the user can redirect the agent mid-task. These contrasting curve shapes make visible the different coordination strategies each product employs.

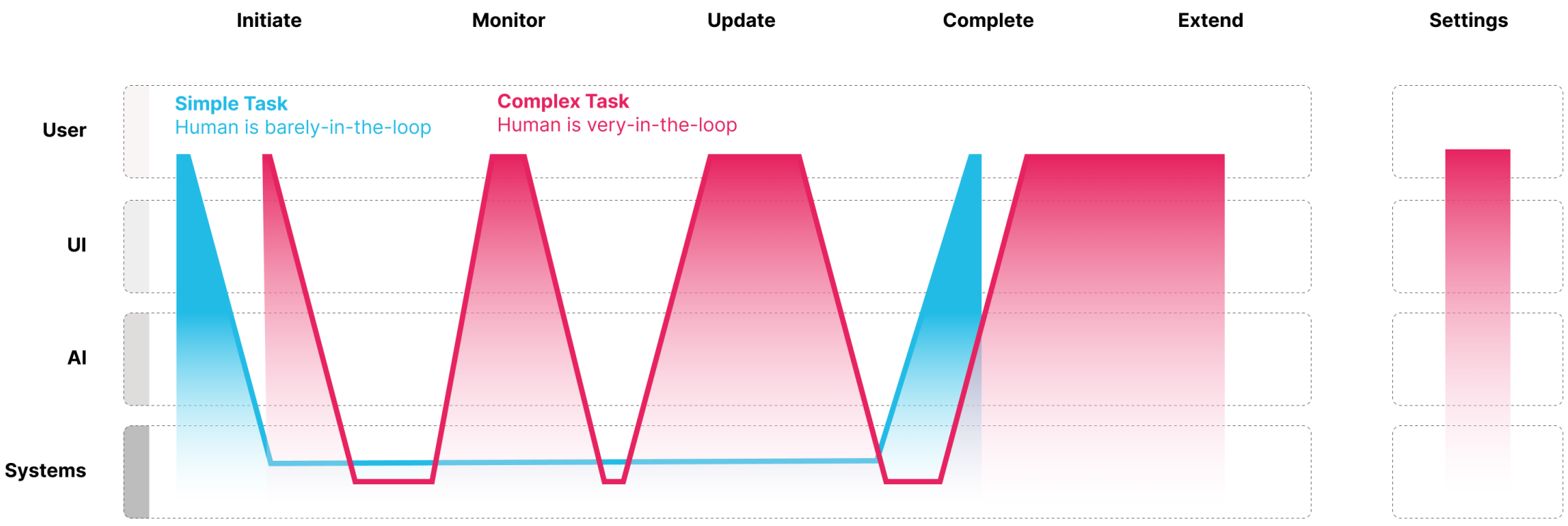

Figure 6. Simplified coordination curves showing contrasting high and low complexity coordinated flows between user and AI.

We introduce two types of coordination curves in our design and analysis work: (1) simplified curves showing the top-level contour of an experience (Figure 6) and (2) granular, annotated curves describing specific UI elements, user touchpoints, and technical components (Figure 7). Both build upon established design tools—including user journey maps and the 5Es framework (entice, enter, engage, exit, extend) [2]—but are adapted to human-AI coordination.

When a specific curve is identified as a useful recurring pattern, we label it a *typology*.[3] Typologies are *abstracted temporal design patterns* (Figure 8). The following sections present coordination typologies drawn from commercially available AI products, demonstrating how our framework vocabulary applies to real-world designs.
While many real-world experiences combine multiple zones, the typologies that follow focus on single-zone coordination patterns to help illustrate the distinct character of each and isolate some of the basic building blocks of more complex composite experiences.

---

[2] Journey mapping [48]is a method used by designers and marketers to visualize a user's or customer's experience with a products, service or brand over time. A layered timeline visualization showing segmented phases and touchpoints is commonly used. The 5Es framework is often combined with journey mapping to design or analyze a user experience across five typical phases of their journey: enticement – what draws them into the experience, entrance – their first interaction with a product or brand, engagement – the main phase of interaction, exit – completion of their main task or activity, and extension – what happens after they leave.

[3] Architects and urban planners, for example, rely heavily on building typologies as a starting point for their designs. These typologies define recurring patterns for the function, form, context, and style of a building. For instance, Le Corbusier famously defined five notable building typologies that he employed, and advocated for. Modern day architects often employ these techniques by drawing diagrams of reusable plans, elevations, and axonometric drawings. The typologies function as building blocks that can be applied, combined, or adapted in a new design. Whereas architecture prioritizes structures, UX design integrally involves *time*. Our coordination typologies focus on this temporal dimension of how coordination various over time.

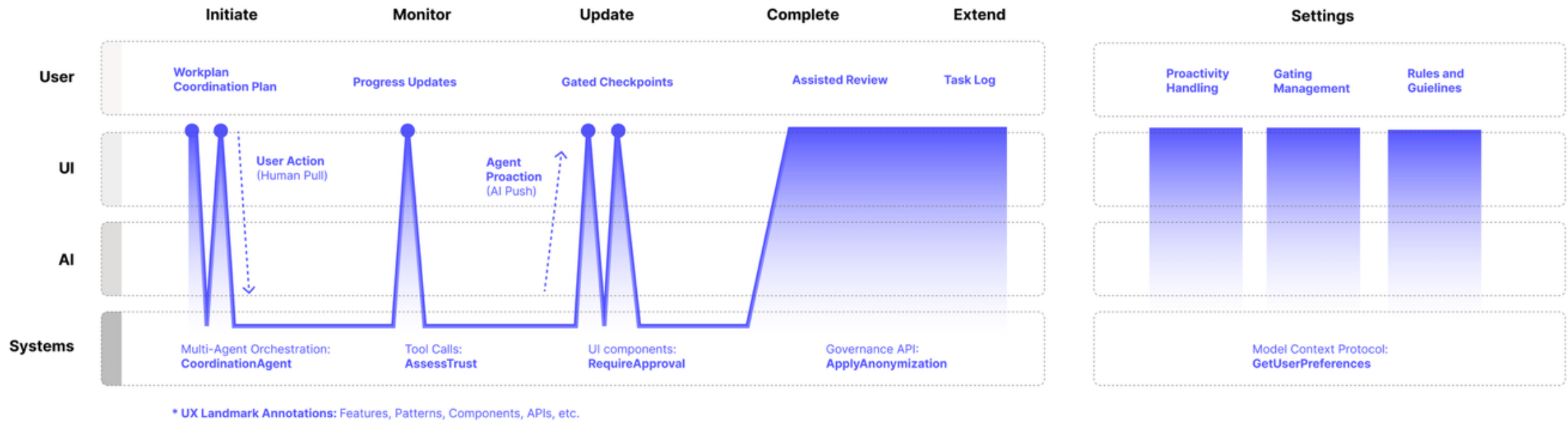

Figure 7. Annotated coordination curves.

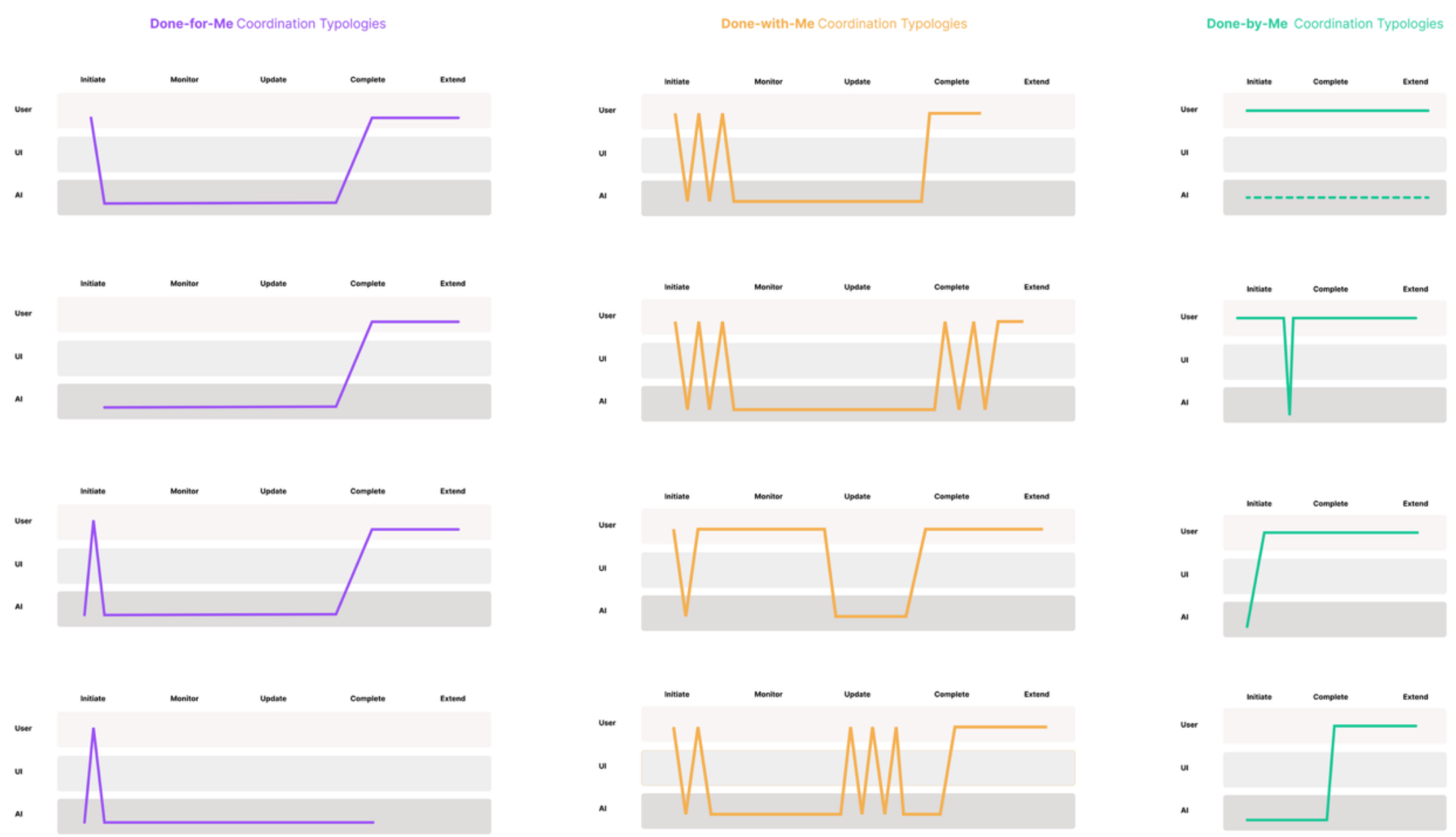

Figure 8: Examples of coordination typologies we extracted from analysis of AI/agentic software products.

#### *4.5.1 Done-For-Me Typologies*

Done-for-me experiences are heavily automated, with user involvement that is typically limited to initiation and/or comple tion phases. We present three done-for-me typologies illustrating varying levels of AI salience.

**Approve and Delegate: Workout Auto-Detection for Apple Watch OS**

Apple Watch OS uses AI to detect exercise activities. When exercise is inferred, the system prompts users with notifications such as “It looks like you’re working out” and a button to accept or reject workout logging. The coordination curve (Figure 8) shows AI working invisibly in the background, with a single brief spike of salience when approval is requested.

Analyzed using our framework, the AI proactively initiates through *inferred input* and *imposed salience* (push notification). The user responds via a *do-it* interaction (accept/reject). This brief moment of salience ensures accurate logging while maintaining low user effort. This auto-detection feature is enabled by default but can be disabled in settings—and an example of pre-initiation that is set by default rather the explicit user consent and choice. However, given that low stakes and high accuracy rate, such a coordination scheme works well across users, as demonstrated by its commercial success.

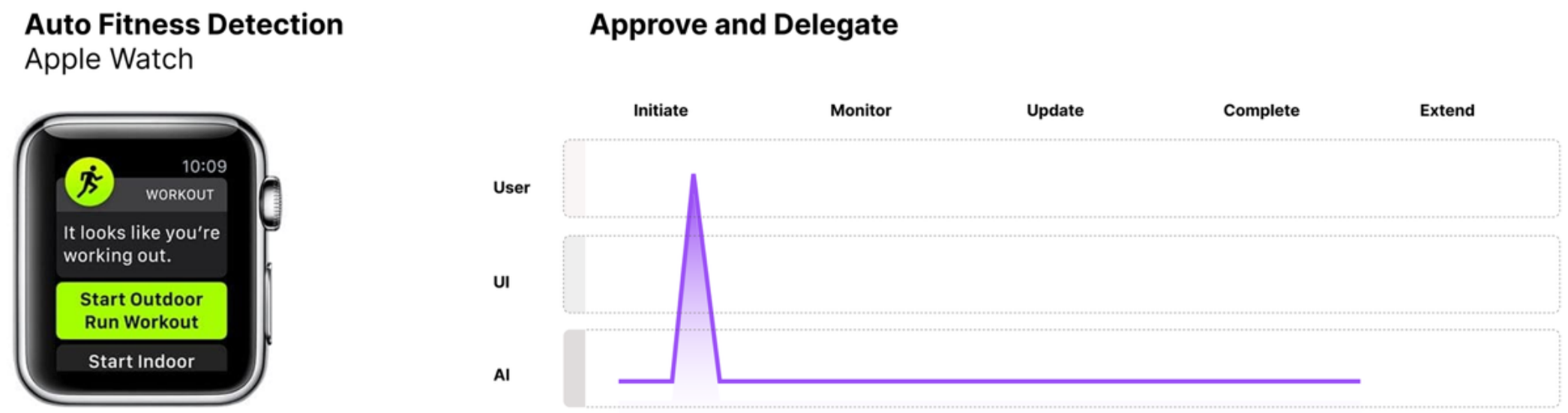

Figure 8: The Approve and Delegate coordination typology, exemplified by the Apple Watch auto-fitness detection feature.

**Pre-Approved Delegation: Gmail Automatic Calendar Entries**

Google Workspace automatically extracts events from Gmail—such as flights, appointments, and dinner reservations—and adds them to the user’s Google Calendar without explicit approval or request. Unlike the Apple Watch pattern, this removes the approval gate entirely. The AI is pre-approved to fully act and complete the task.

The coordination curve (Figure 9) shows minimal salience throughout. Users may eventually notice the calendar entry, but the salient output is subtle and easily missed. This is not a done-without-me experience because users can observe outcomes, but it approaches that boundary. To increase salience, an attribution marker could indicate that “This event was automatically added from your email” and provide users with options to opt-out or adjust this functionality.

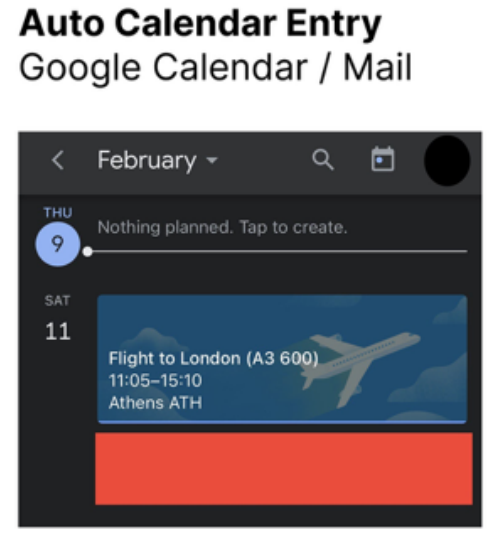

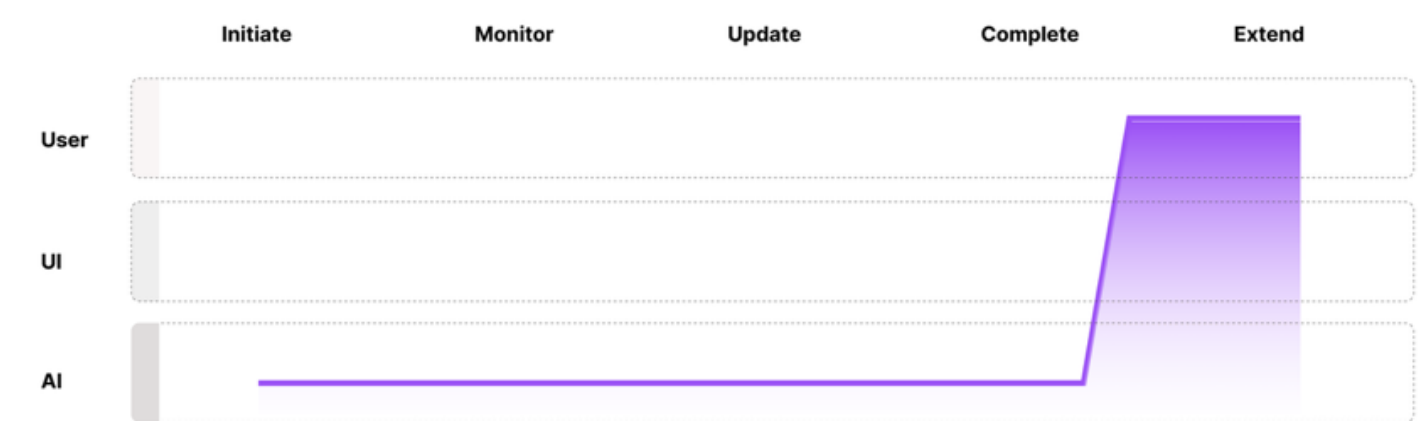

Figure 9. The Pre-Approved Delegation typology, exemplified by auto-calendar entries in Google Gmail and Calendar apps.

**Guided Initiation: Amazon Customer Service**

Amazon's AI customer service chat offers users discrete response options—such as "Package never arrived," and "Thanks, I'm all set"—rather than requiring open text prompts. This *do-that* input pattern reduces initiation effort while maintaining user direction.

The coordination curve (Figure 10) shows streamlined initiation followed by fully automated completion. A user can resolve a missing package with minimal involvement: select an option, confirm details, and receive a replacement item. The experience is *done-for-me* with guided, low-effort initiation and minimal back-and-forth.

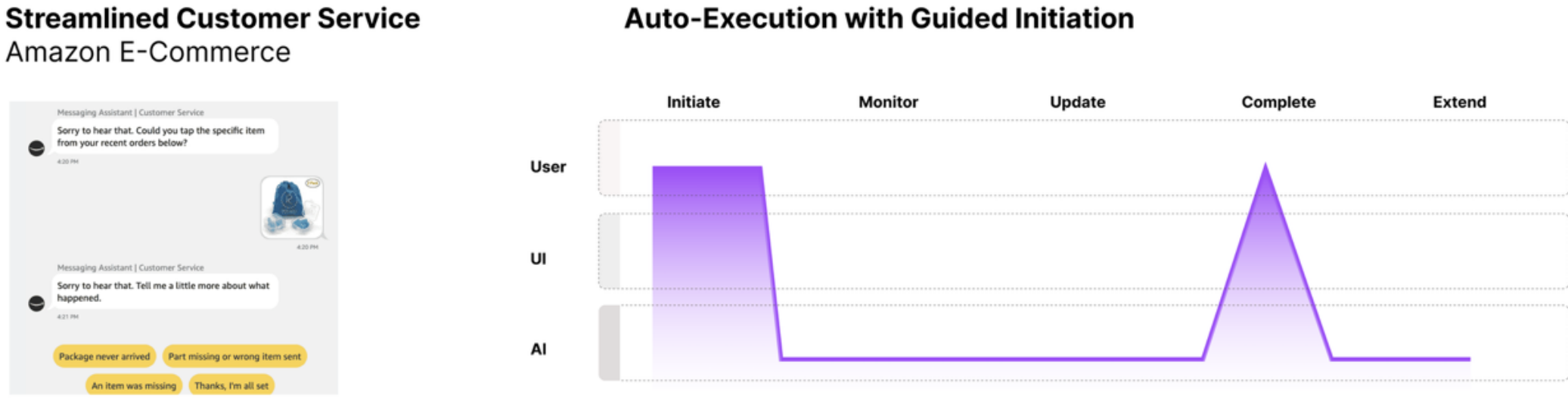

Figure 10. The Auto-Execution with Guided Initiation coordination typology, exemplified by Amazon E-Commerce AI Chat.

*4.5.2 Done-Under-Me Typologies*

Done-under-me experiences involve quiet AI assistance with low or absent salience. Users may not consciously register that AI is helping them. We present two typologies.

**Quiet Nav Inference: Netflix Recommendations**

Netflix uses AI to dynamically curate content based on user behavior and preferences. Video collections appear as navigation cards that users browse to select from. While users feel they are independently choosing which content to watch, AI is quietly presenting titles they are more likely to enjoy.

The coordination curve (Figure 11) shows background AI activity with no explicit salience spikes. Within the user interface, this assistance may be labeled ("Recommended for you") or unmarked. Because salience is minimal, users rarely acknowledge—or even suspect—AI's role in curation. The experience feels entirely conventional, while it is significantly shaped by AI below the interface surface.

**Curated Video Streaming**
Netflix

**Quiet Nav Menu Inference**

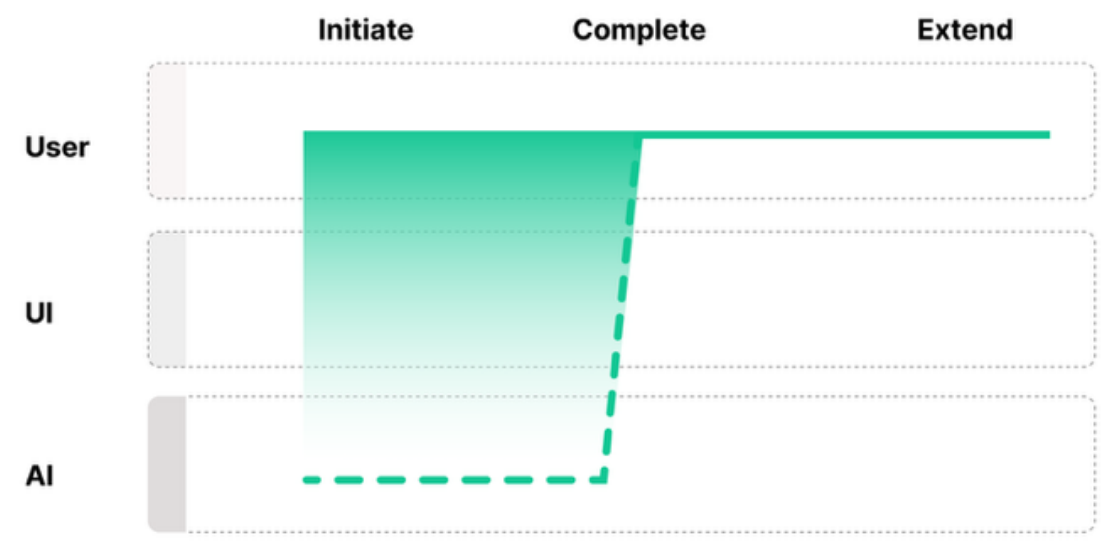

Figure 11. The Quiet Nav Menu Inference coordination typology, exemplified by Netflix AI curated menu content.

**Sparked AI: Google Search AI Overview**

Conventional Search increasingly integrates AI such that certain queries *spark* generative responses. When users enter question-like queries, AI-generated overviews appear above traditional results. Users can selectively extend these outputs into a chat interface for deeper collaboration and deliberate prompting.

The coordination curve (Figure 12) shows conventional search interaction that sparks unexpected AI output. The initial experience aligns with the done-under-me zone: familiar UI produces AI-assisted results without an explicit AI input canvas. Yet with time, users learn that certain query formats (like long questions) are more likely to trigger salient generative AI response, while others (like product searches) yield traditional hyperlink results. This builds on Google's previous Quick Answer features. This pattern illustrates an evolution from purely user-driven search to AI-augmented interaction through *sparked input*.

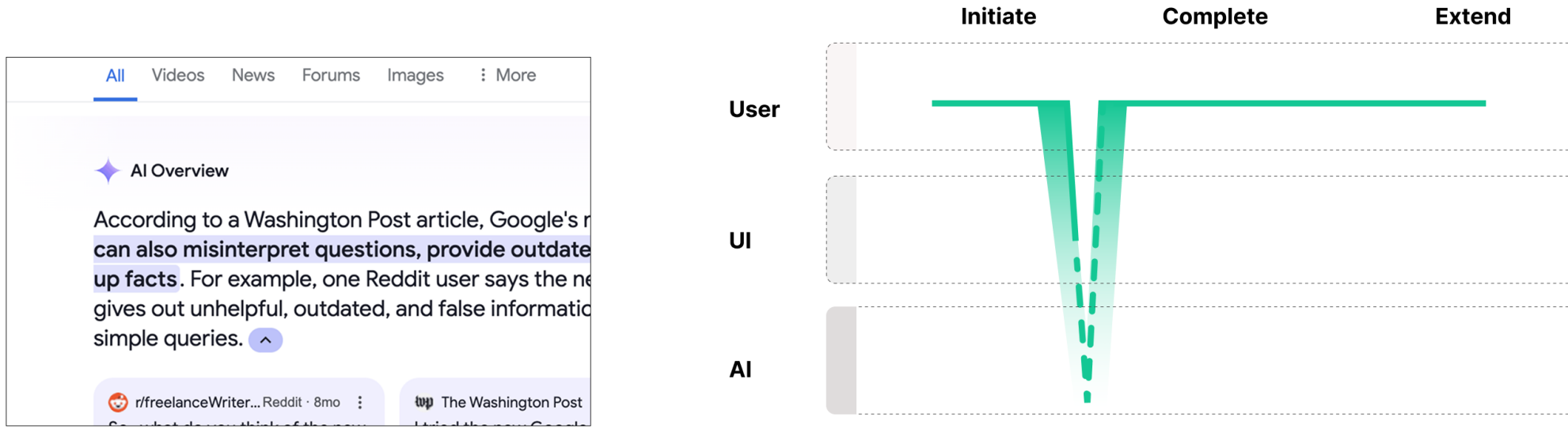

Figure 12. The Sparked Search coordination typology, exemplified by Google AI Search Overview.

### *4.5.3 Done-With-Me Typologies*

Done-with-me experiences involve sustained collaboration across multiple phases, with moderate to high salience and significant user involvement throughout.

**Collaborative workplans: Deep Research versus Manus**

OpenAI's Deep Research and Manus AI both support collaborative done-with-me experiences, but with different coordination profiles. Deep Research allows user to define complex research tasks via detailed prompting (*do-this* initiations). The AI asks clarifying questions, summarizes its plan, then executes. During execution, users can monitor progress with limited indicators but have few intervention options. The coordination curve (Figure 13) shows high involvement at initiation, a valley during execution, and re-engagement at completion.

Manus offers richer collaboration through several UI features: a visual workplan that updates dynamically, a view showing agent activity ("Manus' computer"), and robust support for user interruption and workplan modification mid-execution. The coordination curve (Figure 14) shows multiple peaks of involvement throughout, with iterative revision loops when users interrupt and update the workplan. This enables users to remain consistently in-the-loop across all phases, if they want to be.

The contrast illustrates how done-with-me experiences can vary significantly with respect to *when* and *how* users are able and encouraged to stay involved, even when both qualify as collaborative.

**Deep Research**
OpenAI

**Collaborative Work-Planning, Monitoring, and Review**

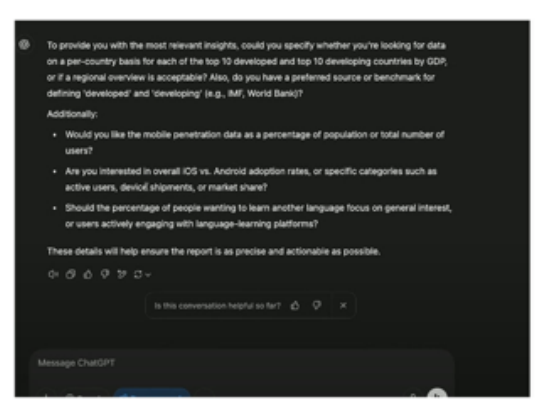

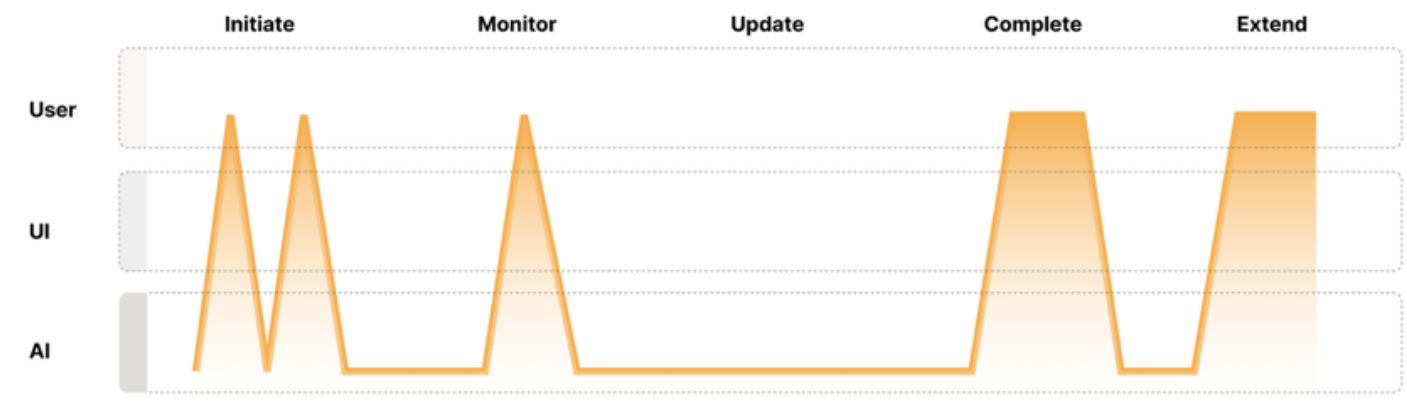

Figure 13. A Collaborative Work-Planning, Monitoring, and Reviewing typology exemplified by OpenAI Deep Research.

**General-Purpose AI Agent**
Manus

**Collaborative Work-Planning, Monitoring, Updating, and Review**

Initiate Monitor Update Complete Extend

User

UI

AI

Figure 14. A Collaborative Work-Planning, Monitoring, Updating, and Review typology, exemplified by Manus.

## 4.6 Comparative Analysis

One additional tool we introduce is a *coordination plot*, which allows quick comparison across products, workflows, or tasks by plotting involvement against salience (Figure 15). We can also visualize AI activity as the size of plotted regions—representing, for example, the duration or cost of AI computation. Such visualizations can help designers identify gaps in their product's coordination coverage and benchmark against existing solutions.

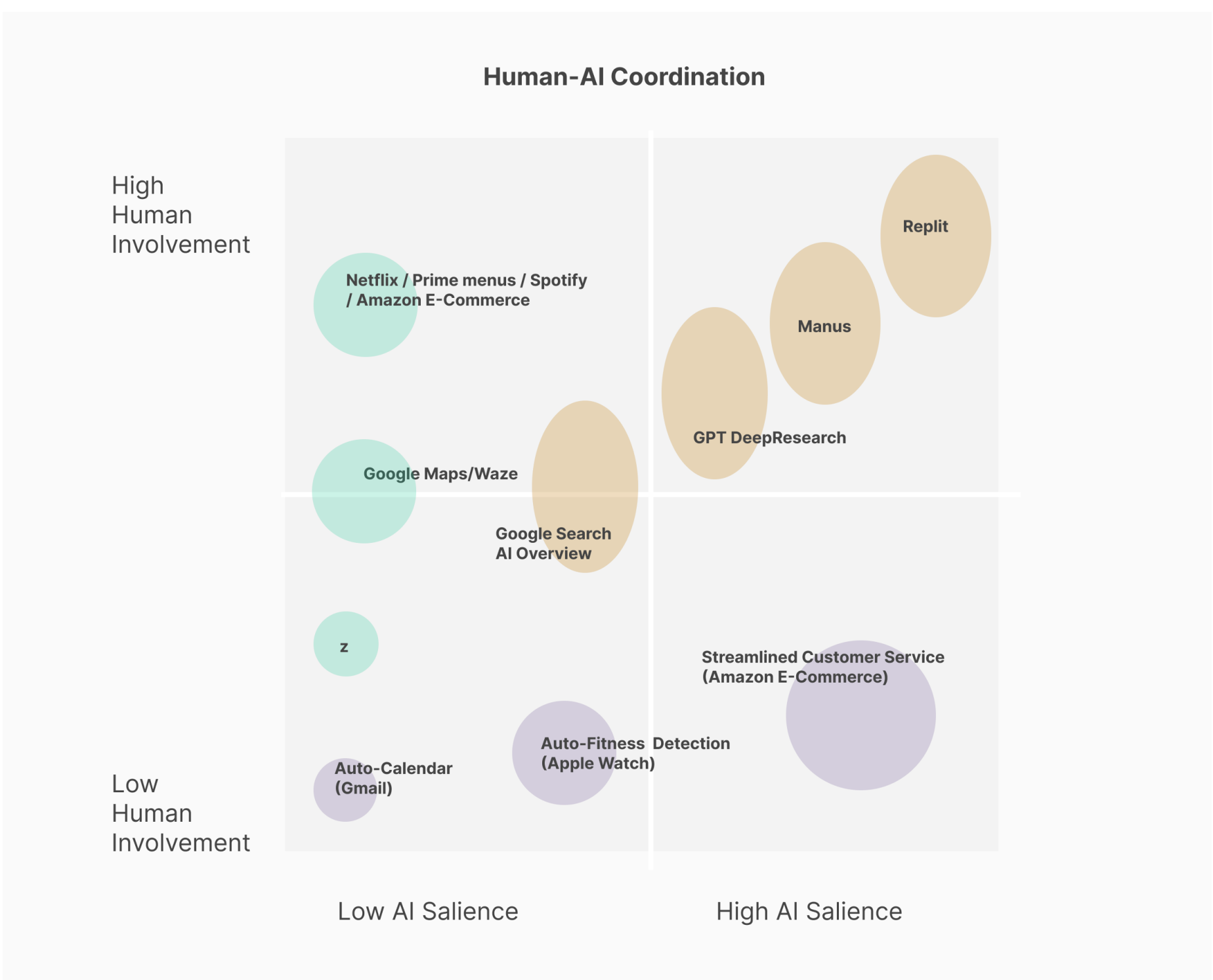

Figure 15: Coordination typologies plotted according to human involvement, AI salience, and AI activity (large size = longer-duration AI activity).

## COMPARISON TO PRIOR WORK ON AI COORDINATION

Having presented our framework, we now return to established concepts in human-AI interaction, human factors, and related fields. We discuss five clusters of prior work: levels of automation, human-in-the-loop, explainable AI and transparency, agency and proactivity, and coordination in human-agent systems. For each, we summarize prior work, show it how relates to our framework, and articulate key differences and points of integration.

### 4.7 Levels of Automation

The concept of varying levels of automation (LOA) is well-established in human factors research. Sheridan and Verplank proposed a foundational taxonomy consisting of ten levels ranging from fully manual to full autonomous [109]. Parasuraman, Sheridan, and Wickens extended this to distinguish four types of activity that can be automated: information acquisition, information analysis, decision and selection, and action implementation [95]. Numerous additional taxonomies have been developed and synthesized in review papers [91] [122]. LOA thinking has also been applied to specific

domains—notably, in the SAE International's six levels of driving automation, an active standard used in the automotive industry [92].

Conceptualizing automation as a graduated spectrum is an intuitive, powerful, and recurring conceptual move. For example, this framing has been applied in HCI and explainability research. Lai and Tan [58], for example, arrange explainable AI techniques along a spectrum of human agency that range from showing only explanations (full human agency) to showing machine predicted labels and suggesting high accuracy (full automation).

Our framework adapts LOA thinking as a flexible, general-purpose tool for considering zones of coordination. However, our framework breaks with the single-dimensional framing (automated versus manual) by introducing two dimensions: salience and involvement.

LOA frameworks focus primarily on *what the system does*—the division of labor between human and machine—rather than *how automation is presented and experienced by users*. Our framework extends this tradition by foregrounding the interaction between AI salience and human involvement: two systems with identical automation levels may present very differently to users, with significant implications for trust, control, and usability.

### 4.8 Human-in-the-Loop (HITL)

Human-in-the-loop (HITL) approaches emphasize keeping humans involved in AI decision-making, particularly for high-stakes applications [18][22][130]. This concept has been influential in machine learning and AI safety research, and it creates a productive contrast with "fully automated" systems. However, HITL is often treated as a binary consideration or intervention in full automation, where the human is either "in" or "out" of the loop. This framing obscures significant variation in *how* humans remain involved, *when* they engage, and *whether they are aware* of their role.

Our framework addresses a complementary question: not whether humans are involved, but how involvement and salience are shaped through the interface across an experience. Coordination zones allow this to be framed as a spectrum, from humans be "barely in-the-loop" (done-for-me), to "unawarely in-the-loop" (done-under-me), to "very in-the-loop" (done-with-me). Beyond the coordination zones, our framework outlines general mechanisms of bringing humans in the loop. Input modes articulate how users get involved through deliberate *prompting sparked* responses that can take users by surprise, or *inferred* input extracted from user activity. A user "in-the-loop" via detailed prompting (*do-this*) is quite different from one who simply approves an AI-generated plan (*do-it*).

The concept of initiation involvement captures the spectrum from higher-effort detailed prompts to lower-effort binary approval to pre-approved autonomous action. Each of these may be labeled "human-in-the-loop" yet represent dramatically different user experiences. Our framework further highlights temporal variation to show that being "in the loop" is not static. A user may be highly involved during initiation, absent during execution, and re-engaged at completion. Our coordination curves attend to these shifts.

Whereas HITL asks *whether* humans are involved, our framework asks *how, when, how much, and with what level of awareness*.

### 4.9 Explainable AI (XAI) and Transparency

Explainable AI research aims to make the reasoning and technical operation of AI systems transparent to users and stakeholders [21][38][80]. HCI has contributed significantly by examining how to design effective explanations [61], when explanations help or hinder users [5], and how developing social-informed definitions [22]. Closely related, research on algorithmic transparency examines user understanding of underlying system operations. Users' lack of awareness and flawed mental models are well documented [24]. Yet research also finds that too much transparency can backfire by reducing trust [42], and that trust can manipulated by, for example, using superfluous or irrelevant explanations [58].

Our framework positions salience as a complementary category that relates to explainability and transparency while foregrounding the experiential dimension—whether users actually attend to, perceive, and act on AI-related information. However, salience hinges on actual user experience: whether AI activity and direction is attended to, consciously perceived, and acted upon. The same explanation or transparency mechanism can have vastly different salience depending on placement, timing, prominence, and user familiarity.

More fundamentally, salience captures whether users experience a system as AI at all. XAI and transparency research often assumes users *know* they are interacting with AI. Our framework addresses cases where this assumption fails. Done-under-me experiences involve quiet AI assistance users may never consciously register. Here, explainability may be irrelevant, as users do not perceive AI presence to begin with—and this may be for the best. Inferred input extracts information from user activity invisibly. Unlike prompted input, where users deliberately engage an AI canvas, inferred input operates with low or no salience. Similarly, sparking input triggers less predictable responses from conventional-seeming UI. Both sparking and inferring represent important, sometimes neglected, sites where designers should consider what level of explanation/transparency may be needed (or not), and how to initiate it.

Our framework also distinguishes how salience is initiated: sought salience (user pulls) versus imposed salience (AI pushes). An explanation users actively request differs experientially from one that interrupts their workflow. Deciding if and how to use generative AI to create and responsively deliver explanations is a significant challenge and opportunity.

### 4.10 Agency, Proactivity, and Mixed-Initiative Interaction

Numerous works addresses the balance between AI automation and human control. Horvitz's principles of mixed-initiative interaction describe how systems can share initiative with users [44]. Related concepts include AI as teammate [106] and human-AI collaboration [127]. Many have emphasized the importance of keeping humans in control [110].

AI *proactivity* is central to these discussions. Yorke-Smith et al define proactivity as the ability to anticipate needs of users, opportunities, and problems, and then take initiative to address them [136].

Proactive systems raise questions about appropriate AI agency: when does AI initiative help users, and when does it overstep? Across this literature, the relationship between human and AI agency is foregrounded. Debates often center on how much agency AI should be granted and how AI initiative may conflict with human autonomy, increase risks, and create harms.

Like explainability and transparency, our concept of salience interacts with proactivity and agency—but shifts focus to the experiential *attention to and perception of* these qualities. Our framework helps designers calibrate and deliver these qualities in relation to an intended coordination zone. For example, high proactivity may be appropriate for collaborative done-with-me experiences but must be handled more carefully with autonomous done-for-me experiences to maintain trust and reduce risks. With done-under-me experiences, it may be best to minimize or constrain proactivity—and the unpredictability and risks it introduces—to achieve the goal of seamless, quiet assistance that users barely notice.

Our concept of salience initiation further operationalizes proactivity as a spectrum between imposed salience (AI push) and sought salience (human pull). With imposed salience, AI proactively surfaces in the interfaces through notifications, suggestions, explanations, and interventions. With sought salience, users initiate engagement through the interface. Here, the AI waits for user action rather than taking initiative.

Agency and proactivity describe system capabilities. Our framework addresses how these capabilities are experienced and mediated through the interface. Two equally proactive systems may feel very different depending on how proactivity is presented (e.g. prominently or subtly, helpfully or intrusively).

### 4.11 Coordination

Coordination has long been studied as human-to-human coordination in HCI and CSCW [74]. More recently, this concept has been applied to human-AI interaction. Klein et al outline ten challenges for human-agent activity [54]. They observe that in human contexts, effective coordination involves mutual predictability (reasonable sense of what others will do), directability (reasonable responsiveness to others), and common ground (shared knowledge, beliefs, assumptions).

Christoffersen and Woods reframe debates about human-automation failure as an issue of coordination [17]. They argue that debates are often incorrectly framed as blaming either human error or over-automation when coordination breakdown is the real culprit. To resolve this, they propose two fundamental design characteristics: *observability* (each party can see relevant aspects of the others' activity) and *directability* (each party can influence the others' activity). Johnson et al extended this by defining "coactive design," and adding *predictability*—each party can anticipate the others' action [50].

Observability and directability roughly map to our definitions of salience and involvement. Salience addresses what users can perceive about AI, and involvement address what users can do to engage or direct it. Our framework translates human factors concepts to user interface design and operationalizes them for UX practice. We add a third dimension—AI activity—which captures what the system actually does, regardless of user perception and action. From a user experience perspective, this highlights the potential mismatches between user mental models and actual system behavior, which is a critical factor in trust and users' subjective *feelings* of control and agency.

Our framework deliberately streamlines coordination to three dimensions. We exclude predictability, common ground, and other dimensions for two reasons. The first is *designer control*. We focus on attributes UX designers can *most directly* seek to manipulate by controlling what is displayed (salience), what actions users can perform (involvement), and what capabilities are executed (activity). Second, our framework treats other dimensions as *emergent qualities*. We suggest that qualities such as predictability, explainability, transparency, and proactivity can be productively understood as arising from the interplay of salience, involvement, and activity—rather than as fully independent design variables. For example, a system becomes predictable when users can observe its behavior (salience) and learn how their actions (involvement) relate to outcomes—and this is all enforced through underlying technical operation (activity).

This streamlined model is intended to facilitate practical design decisions to creatively envision and structure richer coordination qualities that emerge. By foregrounding these three controllable "levers of coordination" at the interface and systems levels, the framework also acknowledges that the list of relevant coordination qualities is inexhaustible.

## 5 DISCUSSION

Our research surfaces three core challenges for crafting AI systems that support useful, usable, and trustworthy human-AI coordination: (1) Defining—how should we define the right type, level, and form of coordination? (2) Designing— if we can define the right level of coordination, how do we iteratively design and build to make it happen? (3) Delivering—how do we construct systems that ultimately deliver the right type of coordination, given that AI and agentic systems are highly non-deterministic and unpredictable compared to conventional interactive technologies? We discuss each challenge, how our framework addresses it, and what new opportunities and challenges emerge.

### 5.1 Defining: Coordination as an Integrative Design Vocabulary

A central contribution of this work is addressing the gap between high-level AI design principles and low-level UI implementation patterns. Existing resources offer either abstract guidance ("be transparent," "maintain user control") or specific component or navigation patterns (chat interfaces, confidence indicators).

Consider progressive disclosure, a mid-level concept implemented through specific UI patterns like accordion menus and tooltips. Progressive disclosure was popularized by Neilsen [87], grounded in insights by Carroll, Rosson, and Carithers[14][15], and continues to be extended in HCI by Springer and Whittaker [114][115].

Our framework can productively extend progressive disclosure through the concept of *progressive autonomy.* With agentic AI, a new design challenge has emerged: users often want tighter control early—while learning the system's limits and calibrating trust—but prefer to reduce their involvement over time. Our framework helps design for this. Experiences can transition from collaborative done-with-me (high involvement, high salience) toward done-for-me or even done-under-me as trust is earned. For example, an AI scheduling assistant might initially require explicit prompted approval for each meeting (done-with-me), then shift to approving batched recommendations through sparked input (done-for-me), and then eventually handle routine scheduling invisibly through purely inferred input (done-under-me). Work begins as a collaborative workplan typology but evolves toward pre-approved delegation.

In general, our coordination zones, curves, and input types function similarly. They are abstract enough to apply across contexts yet concrete enough to guide implementation.

**Why is "coordination" a useful conceptual frame**? Existing concepts—like human-in-the-loop, explainability, transparency, autonomy, and proactivity—each capture important aspects of human-AI interaction. Yet they tend to address different facets of the design challenge in isolation. Our framework proposes three dimensions—salience, involvement, and activity—as integrative levers that can help organize and relate disparate concepts, from autonomy to trust, within a shared design vocabulary.

Consider how salience (the available or experienced perception by a user of AI) relates to several recurring themes in AI design. Concepts related to *understanding* and *observability*—including explainability, transparency, interpretability, intelligibility, legibility—all describe different facets of what users can or do perceive about AI systems. Similarly, concepts related to *initiative* and *timing*—such as proactivity, mixed-initiative interaction, responsiveness, and context awareness—describe how AI presents itself temporally. Our distinction between sought salience (human pull) and imposed salience (AI push) helps describe and operationalize these qualities for design decisions.

Involvement similarly relates to concepts of *control, agency, and power*—like reversibility, controllability, overridability, configurability, user agency. These emphasize what users can do to engage or direct AI systems. Our input types (prompted, sparked, inferred, layered) and initiation involvement spectrum (do-this, do-that, do-it, already-done) provide generalized mechanisms for supporting and inviting user involvement.

Our framework focuses on attributes UX designers can more directly craft: What is displayed (salience), what actions users can perform (involvement), and what capabilities are executed (activity).

We suggest that other important qualities—such as predictability, trustworthiness, and proactivity—are shaped in part by the interplay of salience, involvement, and activity. For example, a system may become more predictable when users can observe its behavior (salience leading to observability) and learn how their actions (involvement tied to directability) are related to outcomes.

### 5.2 Designing: Mid-Level Tools for Human-AI Coordination

As discussed in the Introduction, practitioners lack mid-level design knowledge bridging high-level principles and low-level patterns [134]. Our framework responds to this gap in several ways.

Our coordination zones help practitioners frame design situations. Rather than abstractly asking "how transparent should this be?" designers can ask "is this task or workflow best offered as a done-for-me, done-with-me, or done-under-me experience?" This reframing surfaces concrete design implications. For example, done-for-me experiences concentrate involvement and salience at initiation and completion phases. Done with me experiences require sustained collaboration mechanisms. Done-under-me experiences minimize salience to achieve seamless assistance, but they must be implemented carefully to maintain trust; they are best when AI output is quick, clear, and quiet.

Our coordination curves adapt established journey mapping techniques for AI contexts. Unlike conventional interactive systems, AI-enabled systems may include significant periods where users are uninvolved, and yet AI is conducting sophisticated and potentially costly and risky work. These periods may require notification, approval, monitoring, or correction. These "valleys" in the coordination curve are just as important as the "peaks," which represent moments of higher salience and involvement.

In our own professional design work, we have applied our framework to synthesize clear and actionable design patterns for coordinating human-agent work. While some of these patterns may demonstrate novelty (e.g., responsive salience), we present them here as a real-world demonstration of how our framework can be applied generatively to *identify and articulate* such patterns. Assisted by our framework, we have developed the following design patterns and used them to design, build, and evaluate functional agentic applications:

- **Responsive salience**: Dynamically adjusting AI prominence and involvement options based on context, user behavior, or task characteristics. A system might increase salience attributes in the UI—such as chain of thought transparency or next-step proactivity—when trust is low or stakes are high, and decrease it when operating reliably on routine tasks. Inferred input or context prompts can be used to intelligently assess contextual attributes like trust, risk, expertise, and emotional state.
- **Workplan Gating:** Specifying when, where, and how users will coordinate with AI during a workflow. Gating is especially useful when applied to a pre-determined workplan. Gates are points in a workplan where users assign coordination parameters—such as approval requirements, notification preferences, review checkpoints, or escalation rules. Segments are the periods of AI work occurring between gates. Gating helps users plan future coordination rather than simply reacting to AI outputs. They are especially useful for complex, costly, or long duration workflows where users want the hands-off benefits of a done-for-me experiences with the risk mitigation and control benefits of a done-with-me experience.
- **Attribution Markers:** Making the contributions of AI activity visible through annotations, labels, or indicators (e.g., "This review was generated by AI"). Attribution layers can be more or less salient depending on context. Prominent warnings may be appropriate for high-stakes content or policy requirements, while subtle indicators are better for routine workflows.
- **Progressive Autonomy:** Gradually increasing AI independence and autonomy as trust is established, time passes, or permission is granted. A system might begin in done-with-me mode, requiring frequent approval and transparent thinking, then transition toward done-for-me , -under-me, or -without-me as users demonstrate comfort and system demonstrates reliability.

These patterns are examples of how we have used our framework ***generatively*,** to synthesize and communicate new lower-level UX design patterns. This broad generative capacity distinguishes mid-level how-to tools from both abstract what-should principles (which lack implementation guidance or direction) and detailed UI components or patterns (which lack conceptual grounding and extensibility).

### 5.3 Delivering: Managing Variability and Non-Determinism

Agentic AI's core value—its autonomy and decision-making capabilities—is also what makes it a difficult design material: outputs are non-deterministic and cannot be fully predicted or guaranteed [103][128]. Designers must therefore craft coordination in ways that support trust, control, transparency, and other desired qualities, since users cannot rely on the same assurances they have with traditional interfaces.

Our coordination curves help model this variability. We used coordination curves to identify effective typologies for different experiences, such as approve-and-delegate (brief approval within automation), guided initiation (structured entry into automated work), and collaborative workplans (sustained back-and-forth throughout execution). These typologies may help teams design features capable of supporting and dynamically adjusting coordination by altering salience, involvement, and activity.

The design pattern for responsive salience that we introduced above directly addresses delivery challenges. Rather than designing for a single, fixed coordination profile, designers may specify how salience should adapt based on runtime conditions. Generative AI can be employed to assess and triage such conditions by, for example, increasing transparency when AI confidence drops, offering more involvement options when tasks become complex, or reducing interruptions when users demonstrate proficiency. This changes coordination from a static or deterministic branching design decision to a dynamic and responsive system capability.

Our framework also highlights where conventional design methods struggle. Several teams at our organization described how standard methods they rely on for design and development—including user stories, user flows, and UI mockups—did not work as well when designing highly unpredictable AI/agentic systems. User journeys for AI experiences may branch unpredictably or shift between coordination zones mid-journey. Our phases (initiating, monitoring, updating, completing, extending, and settings) and our concept of composite workflows provide a conceptual toolkit for capturing this complexity.

### 5.4 Limitations

One limitation of this work is our reliance on artifact analysis, expert judgment, and literature review rather than extensive empirical user validation. While we validated our framework through internal stakeholder feedback and applications to real design and development projects, additional design exemplars and user studies are needed to further empirically validate the concepts we developed. In support of our approach, however, we note that many influential mid-level design tools were created from practice-based research and expert judgment and were later validated more thoroughly through subsequent uptake by others.

A second limitation is our corpus construction. By excluding most enterprise software, research prototypes, and expensive specialized products, we likely omitted many novel and early-stage examples of AI UX techniques. Our sampling strategy sought to prioritize breadth and variation over exhaustive coverage which allowed us to identify a wide range of coordination patterns. Yet we likely missed emerging and innovative approaches in specialized domains.

Third, our corpus was draw primarily from consumer-facing AI applications. Applying the framework to high-stakes domains—such as healthcare, legal analysis, or safety-critical systems—may expose the need for domain-specific adaptions. In such contexts, the appropriate calibration of salience, involvement, and activity may be shaped by regulatory requirements, professional norms, and the severity of potential harms in ways that our current analysis does not fully address. Our framework is a starting point that provides a useful operational vocabulary, but we recognize that the specific coordination patterns appropriate for a clinical AI assistant, for example, may differ substantially from those of a consume recommendation system. Future work should examine how the framework's constructs translate across domains with different stakes, user expertise, institutional constrains, and corresponding UX/UI products.

### 5.5 Future Work

Our research exposes many avenues for future work. First, there is a **need for AI UX benchmarks, standards, patterns, and exemplars**. It has taken decades for UX to establish standard components and patterns, like cards, tooltips, autocomplete, and guided onboarding tours. In the meantime, practitioners find themselves adrift without clear, mid-level reference points. Presumably such knowledge will eventually emerge for AI. Our framework provides scaffolding for generating and organizing this emerging knowledge base as it develops.

Second, **coordination can be quantified through various metrics.** Human involvement can be quantified with UX metrics such as time-on-task, click rate, or self-reported assessments of effort. Salience may be quantified with self-reported assessments and further broken down in specific types such as user-perceived "autonomy," "proactivity/initiative," and "visibility." Metrics could include click repetition for prominent AI UI elements, or the number of chat dialogs or other forms of human-agent turn-taking. AI activity can be quantified with various metrics of computation, time, and cost (e.g., cost per instance, request throughput, token throughput). Future work could develop techniques for measuring coordination dimensions. This can support both evaluation of their effectiveness in different contexts and be used to inform inferred or sparked input to structure interaction (e.g., through responsive salience).

Third, there is a **need for updated rapid prototyping techniques for agentic AI**. GenAI has opened up exciting new design and prototyping possibilities, along with issues on organizational integration [120]. Yet standard techniques reliant on deterministic input-output relationships and well-defined user tasks often fail in the face of non-deterministic AI systems that respond to user intentions, or whims—what Neilson has described as a new AI design paradigm of "intent-based outcome specification" [84]. While agentic design and development tools have thus far focused on generative UI screens and functional prototypes, there are under-explored opportunities for applying generative AI to the creation of non-UI artifacts, such as journey maps, sitemaps, flow diagrams, and coordination curves. For example, AI might enable dynamic visualizations of coordination curves, in a way that allows designers to abstractly model a user journey, and see both routine experiences and occassional edge cases, or preview the expected distribution of sparked UI outputs for a persona or task.

## 6 CONCLUSION

Addressing a gap between high-level AI design principles and low-level UI patterns, we present a mid-level framework of human-AI coordination. Developed through analysis of 60 commercial AI applications, the framework introduces coordination zones, input types, and coordination curves as concepts for defining, designing, and delivering varying degrees and forms of human-AI coordination. (See Appendix A0 - Table 2 for a summary.)

We think of our mid-level knowledge framework as a foundation for a playbook of concepts, strategies, and tactics for expert practitioners to apply and adapt to diverse AI/agentic design situations. These tools span different levels of abstraction and work together to provide flexible mid-level implementation guidance.

Our framework can be applied in two basic ways. First, it can be applied *generatively* to design challenges by using zones to frame requirements, curves to plan experiences, and input types to structure user engagement and compose interfaces. It can further be used to invent and refine new lower-level implementation guidance, such as our design patterns for responsive salience, gating, attribution layers, and progressive autonomy.

Second, the framework can be applied *analytically* to existing designs to identify and name coordination patterns and typologies, assess tradeoffs, and communicate challenges and opportunities. Third, the framework is a communicative tool for sharing knowledge across stakeholders and communities of practice.

These practical challenges connect to a broader systemic one: **how can we trust and control AI systems that are inherently difficult to trust and control** owing to their indeterminacy [7] and opacity [125]? This paper contributes a framework of concepts and techniques for improving trust and control through the design of well-coordinated human-AI experiences. Like any technology, AI is neither good nor bad; but neither is it neutral [57]. While we have scoped this paper around how-to questions of technique, we urge practitioners and researchers to situate our work within what-should principles and the ethical imperatives of safety, trustworthiness, transparency, and fairness. Values-based design is essential—but it ultimately requires mid-level conceptual tools to translate into practice.

# A APPENDICES

## A.0 Tables

| Types of Done-For-Me experiences | UI Example |
|---|---|
| **Do-this**: The user specifies with unique detail the task for the AI/agent to complete. The most common format for do-this experience is a chat dialog or open-ended text field for the user to input a verbal prompt. In more involved and collaborative done-for-me experiences, the initiation phases may entail extensive back-and-forth interaction resulting in a refined, visible, and user approved work plan. | **What can I help with?**<br><br>Example of a Do-This UI: Open-ended text prompt field for ChatGPT. |
| **Do-that**: The user selects from a set of options presented by the AI/agent. There is less back-and-forth collaboration during the initiation phase. Instead, the agent proactively suggests options and the user selects. | 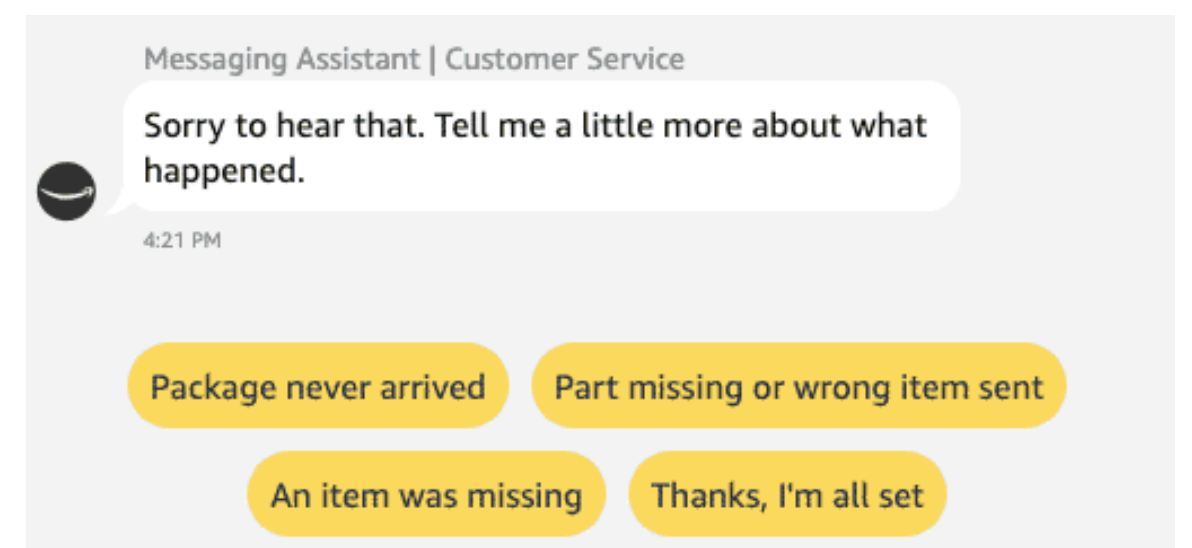<br>Example of a Do-That UI: Multiple AI-generated response buttons, including “Package never arrived” and “Thanks, I’m all set” offered during Amazon’s e-commerce custom support agent experience. |
| **Do-it:** The user approves a plan or task the AI/agent has prepared and is ready to execute. | 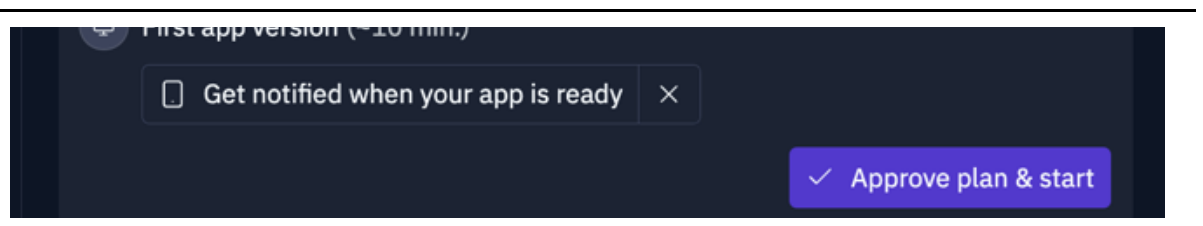<br>Example of a Do-It UI: “Approve plan & start” button in Replit. |
| **Already-done** or **already-doing**: The AI/agent proactively anticipates outcomes the user may desire and presents them without explicit request from the user. The user can then accept as-is, or extend the work into a different type of experience (e.g., collaborative, do this instead, etc.) | 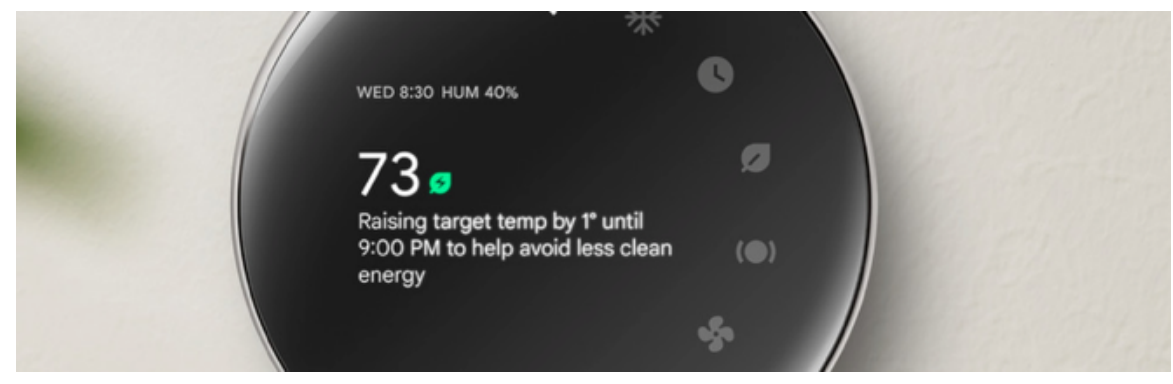 |

| | Example of an Already-Done UI: Nest thermostat can automatically make temperate adjustments based on inferred user preferences. The Energy Shift smart feature, for example, can automatically shift energy usage to cleaner or cheaper times. |
|---|---|

Table 1: Four types of done-for-me experience (from most to least human involvement required).

| Key Concept | Summary |
|---|---|
| **Human-AI Coordination** | The interplay of human involvement, salience of AI, and AI activity. Expressible as a ratio of human involvement to AI salience (low coordination has low involvement and salience; high coordination has high involvement and salience). |
| Involvement (of a user) | The degree and forms of effort a user invests, or is demanded of them, with an AI capability or system via a user interface. |
| Salience (of an AI to user) | The prominence of AI: The extent to which AI is perceived by a user *as AI* via the user interface. AI that is more visible, demanding, vocal, available, transparent , proactive, anthropomorphic and so on will typically have greater salience to users. |
| Activity (of an AI) | The actual backend system activity of an AI, regardless of how much is saliently exposed for the user. |
| **Coordination Zones** | Four types of human-AI coordination, based on the ratio of human involvement to agent salience. |
| Done-for-me | Work is *heavily automated* and completed by AI with little or no user assistance.<br>The human is *barely in-the-loop*. |
| Done-under-me | Work is performed by the user while *quietly assisted* by AI operating in the background. AI salience is low or absent. The human is *unawarely in-the-loop*. |
| Done-with-me | Work is *collaboratively performed* with AI, and the human is significantly involved. The human is *very in-the-loop*. |
| Done-without-me | Work *covertly conducted* by AI without salience or conscious human involvement. The human is *fully out-of-the-loop*. |
| **Input Involvement** | Types of user-involved input into an AI system via the user interface (UI) |
| Conventional input | A user acts upon a conventional UI element (e.g., nav menu, data entry field) and the system reacts in predictable ways (e.g., navigating to a page, confirming data entry). |
| - Prompted input | A user consciously and deliberatively interacts with an *AI input canvas* and which leads to recognizable generative AI outputs such as AI content cards or chat dialogs. |

| Key Concept | Summary |
|---|---|
| - Action prompts | "Do something:" Requests for tasks or outputs, such as answering questions, creating content, or executing commands. |
| - Context prompts | "Know something:" Information the user provides—such as examples, preferences, or background—that do not necessarily request immediate action. |
| - Guide prompts | "Follow something:" Constraints, rules, or boundaries the AI should respect, such as tone, format, or guardrails. |
| Sparked input | User input generates an AI response that qualitatively breaks from expected output patterns, presenting novel or unexpected UI elements. |
| - Cold sparks | Conventional UI interaction (button click, hover, scroll) unexpectedly triggers generative AI output. |
| - Warm sparks | Prompted AI interaction produces a qualitatively different output *typ*e than expected, rather than the typical variability characteristic of prompted outputs. |
| Inferred input | A system intelligently infers or extracts information from a user's activity and uses it to trigger an AI activity. Salience of the inferred input is low or absent. |
| Layered input | Prompting, sparking, and inferencing are layered upon one another and conventional inputs. |
| | |
| **Saliency Initiation** | How and by whom a salient AI action is initiated. |
| Sought saliency (human pull) | AI saliency is made available to a user based on an input request or motivated search. |
| Imposed saliency (AI push) | AI saliency is imposed on the user's environment proactively and without explicit user initiation. |
| | |
| **Initiation Involvement** | |
| Do-this prompting | User specifies the task in detail via open text or speech. |
| Do-that prompting | User selects from AI-generated options. |
| Do-it prompting | User approves or rejects an AI-prepared option. |
| Already-doing/done | AI proactively decides/acts with pre-approved or post-approved user permission. |

Table 2: Summary of conceptual building blocks for defining human-AI coordination.

## A.1 Artifact and Landscape Analysis

### *A.1.1 Corpus: List of AI Products and Features Reviewed*

| | Product Name | Summary Description |
|---|---|---|
| 1. | Perplexity | An AI-powered search engine and chatbot that provides accurate and sourced answers to user queries. |
| 2. | Suno | AI music generation tool that creates songs and instrumentals based on text prompts. |
| 3. | Janitor AI | AI chatbot platform with customizable personalities for entertainment and roleplay. |
| 4. | Quillbot | AI writing assistant that helps with paraphrasing, summarizing, and grammar correction. |
| 5. | Poe | A chatbot aggregator by Quora that allows access to multiple AI models in one place. |
| 6. | Liner | AI-powered web highlighter and research tool that helps users organize and summarize content. |
| 7. | Civitai | A platform for sharing and discovering AI-generated images, models, and LoRAs. |
| 8. | Luma | AI-powered 3D content generation and video enhancement tool. |
| 9. | Leonardo | AI image-generation tool focused on high-quality, creative outputs for artists and designers. |
| 10. | Midjourney | A popular AI image generator that produces detailed and artistic visuals from text prompts. |
| 11. | Gamma | AI-powered tool for creating presentations, documents, and decks with minimal effort. |
| 12. | Pixlr | AI-enhanced online photo editor with powerful editing and design tools. |
| 13. | You.com | A privacy-focused AI search engine with chatbot functionalities and productivity tools. |
| 14. | DeepAI | AI platform offering text-to-image generation, deep learning models, and other AI tools. |
| 15. | SeaArt | AI image-generation tool for creating digital art and illustrations. |
| 16. | InVideo AI | AI-powered video creation and editing platform for generating professional-looking videos. |
| 17. | Udio | AI music-generation tool that creates high-quality compositions from text prompts. |
| 18. | Chatbot App | A general chatbot application with AI-driven conversations and automation features. |
| 19. | Vocal Remover | AI tool that isolates vocals and instruments from audio tracks. |
| 20. | Vidnoz | AI video generator for creating talking avatars and deepfake-style content. |
| 21. | MaxAI | AI-powered assistant with multiple tools for writing, summarization, and productivity. |
| 22. | ChatPDF | AI tool that allows users to chat with and extract insights from PDF documents. |
| 23. | Gauth | AI-powered homework helper and tutoring tool for students. |
| 24. | Coze | AI chatbot platform for businesses and developers to create conversational AI. |
| 25. | Playground | AI-generated art and image tool similar to DALL·E and Midjourney. |
| 26. | Speechify | AI-powered text-to-speech tool that converts written content into natural-sounding audio. |
| 27. | Jasper | AI writing assistant for marketing, content creation, and business communications. |
| 28. | NotebookLM | AI-powered research assistant by Google that helps summarize and analyze documents. |
| 29. | UX Pilot | AI-driven UX research and design assistant for user experience professionals. |
| 30. | Subframe | AI-powered design tool for generating UX/UI layouts and design concepts. |
| 31. | GitHub Copilot | IDE-integrated AI assistant for developers. |
| 32. | Figma AI | Integrated AI assistant for designers and developers. |
| 33. | FlowiseAI | Open-source UI tool designed to help you create and manage custom LLM flows using LangchainJS |
| 34. | StackAI | An enterprise-grade, no-code platform for building, deploying, and scaling AI agents and applications |
| 35. | Microsoft AutoGen Studio | Rapid prototyping for multi-agent workflows. |
| 36. | CrewAI | An open source multiagent orchestration framework. |
| 37. | Relevance AI | Platform for building and deploying an AI workforce. |
| 38. | Cline | Open-source AI coding agent. |
| 39. | Elicit | AI-powered research assistant. |
| 40. | Pippin | AI development platform for agents. |
| 41. | Runway AI | Creative platform and toolset that uses generative AI to help users create, edit, and enhance visual content like videos and images. |
| 42. | LinkedIn | A social networking platform focused on career development and professional networking, with AI features. |
| 43. | Amazon.com | Large e-commerce site features AI customer-service chat, recommendations, and review summaries. |

| 44. | Amazon Prime Video | Media streaming service featuring AI-generated recommendations and content. |
|---|---|---|
| 45. | Netflix | Media streaming service featuring AI-generated recommendations and content |
| 46. | Google Maps | Map and navigation app featuring AI-generated routes, menus, and review summaries. |
| 47. | Adobe Creative Cloud | A collection of creative software applications and online services for photography, graphic design, video editing, web design, and more; recent updates include multiple AI features. |
| 48. | Canva | Online graphic design platform with AI features. |
| 49. | OpenAI Operator | An AI agent, or computer-using agent (CUA), that performs tasks by interacting with web browsers and applications as a human would. |
| 50. | OpenAI Deep Research | An advanced AI agent that conducts complex, multi-step research on the internet to synthesize information from public web pages, uploaded files, and third-party data sources into a documented, cited report |
| 51. | Open AI Chat GPT | An advanced AI chatbot. |
| 52. | Anthropic Claude | An advanced AI assistant. |
| 53. | DeepSeek | An advanced AI assistant. |
| 54. | Google Gemini | An advanced AI assistant. |
| 55. | Manus AI | An autonomous, general-purpose AI agent. |
| 56. | Replit | AI-powered IDE. |
| 57. | Otter.ai | AI meeting transcription and summarization |
| 58. | Zoom AI Companion | Video communication platform features for meeting summaries, transcription, and smart scheduling |
| 59. | Google Photos | AI powered photo search and curation. |
| 60. | Apple Photos | AI powered photo search and curation |

#### *A.1.2 Analysis of AI UI*

Below are samples of our process analyzing existing AI/agentic applications. Our process involved:

- We used AI products (firsthand or through video demos and video reviews) to familiarize ourselves with their UI and functionality.
- We curated screenshots of noteworthy UI.
- We used a simplified heuristic evaluation approach of annotating screenshots with (1) notable UI features and patterns, (2) strengths, (3) weaknesses, (4) and questions/hypotheses (e.g., "too much transparency is overwhelming?").
- We grouped recurring UI patterns, features, and elements into themes (e.g., "AI action bars," "attribution layers").
- Using our coordination curve technique to map user journeys and corresponding UI and AI features.

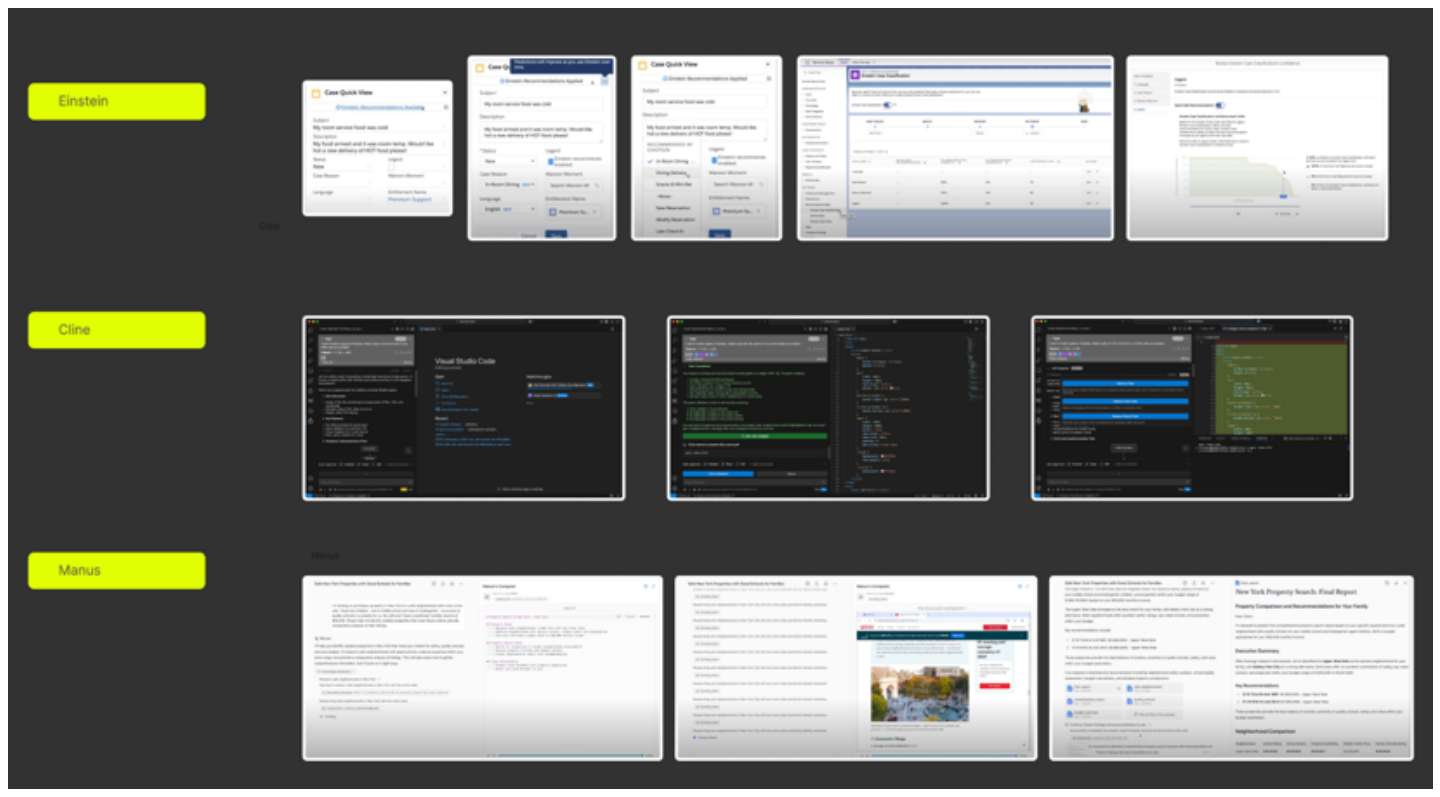

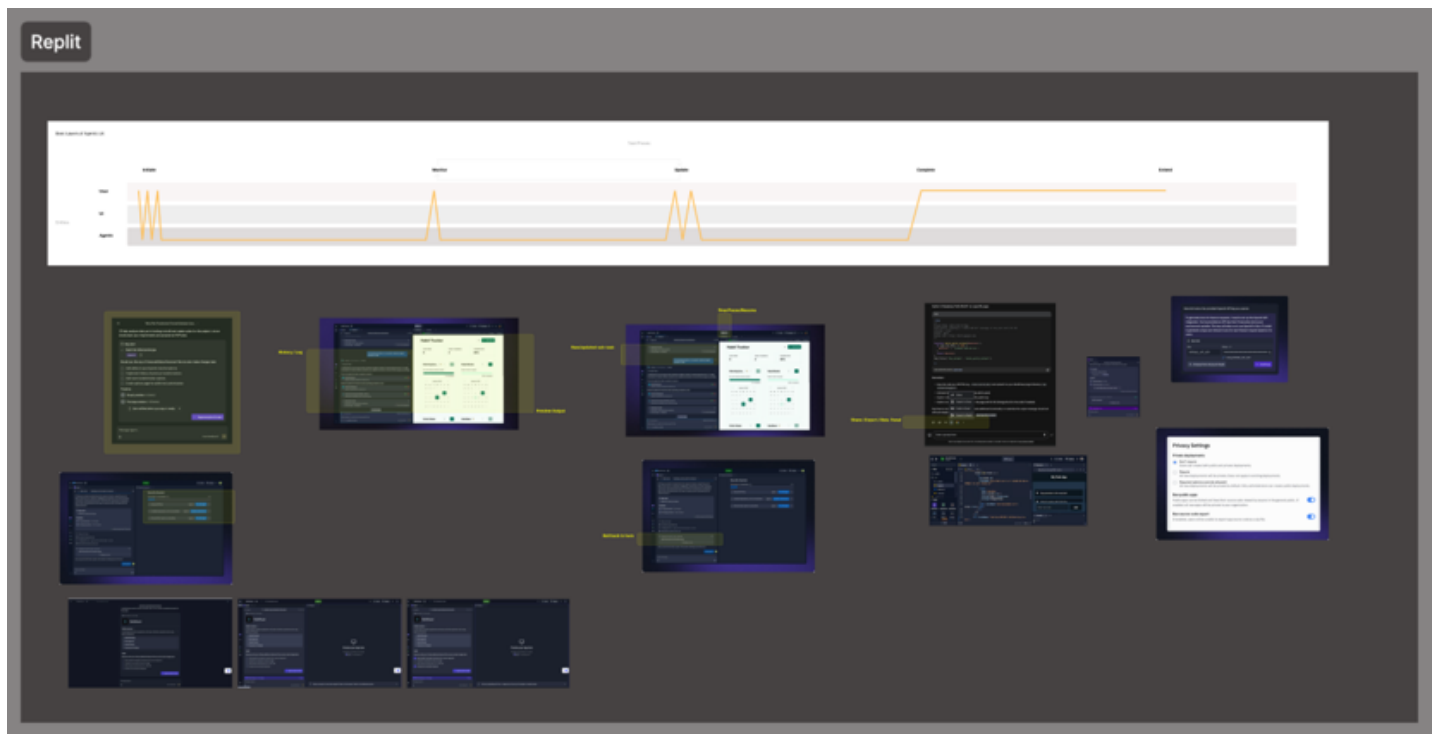

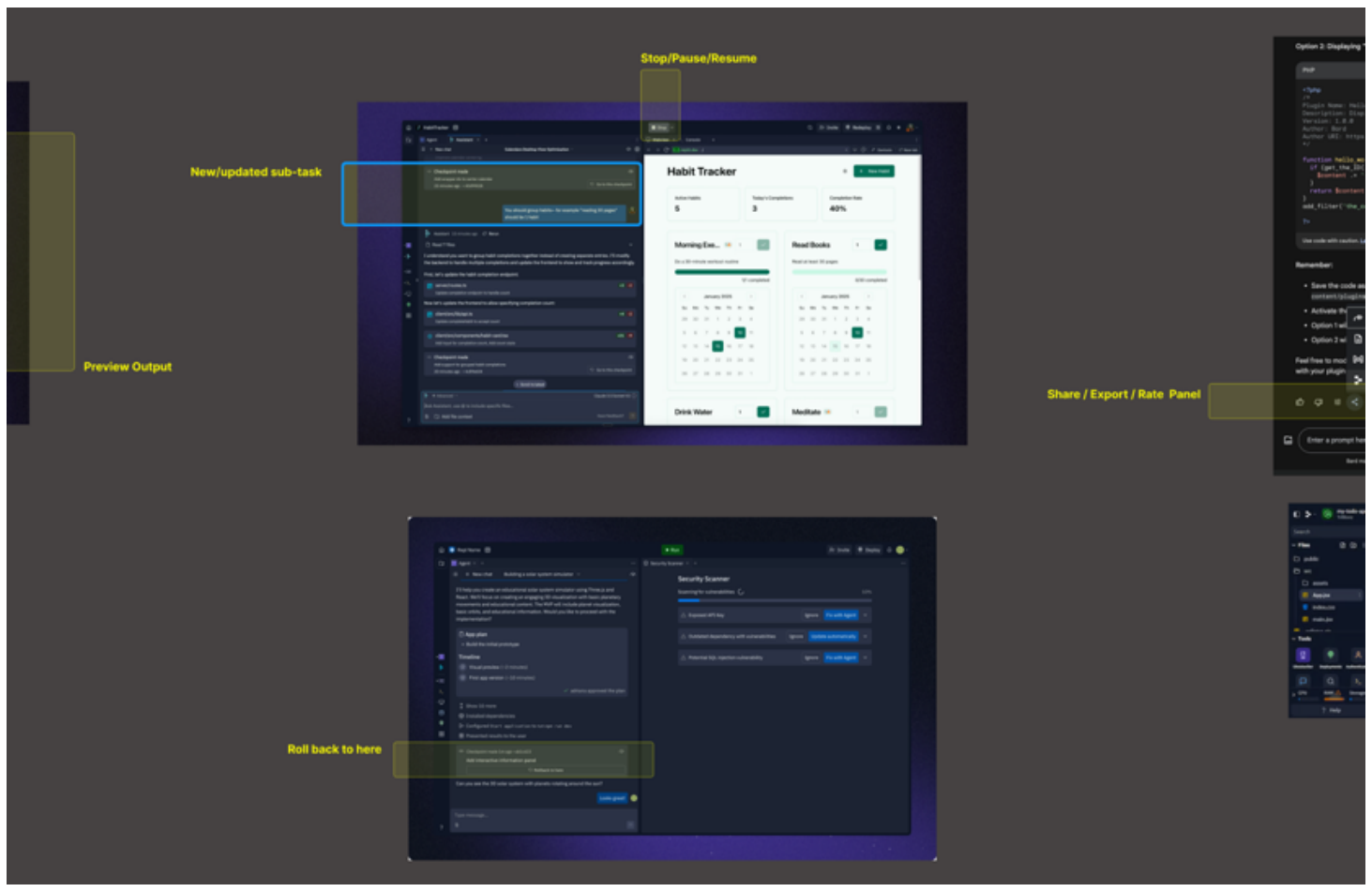

Examples of our analysis of a specific AI application called Replit, an AI-powered software development tool. We annotated screenshots with new and emergent AI-centric features.

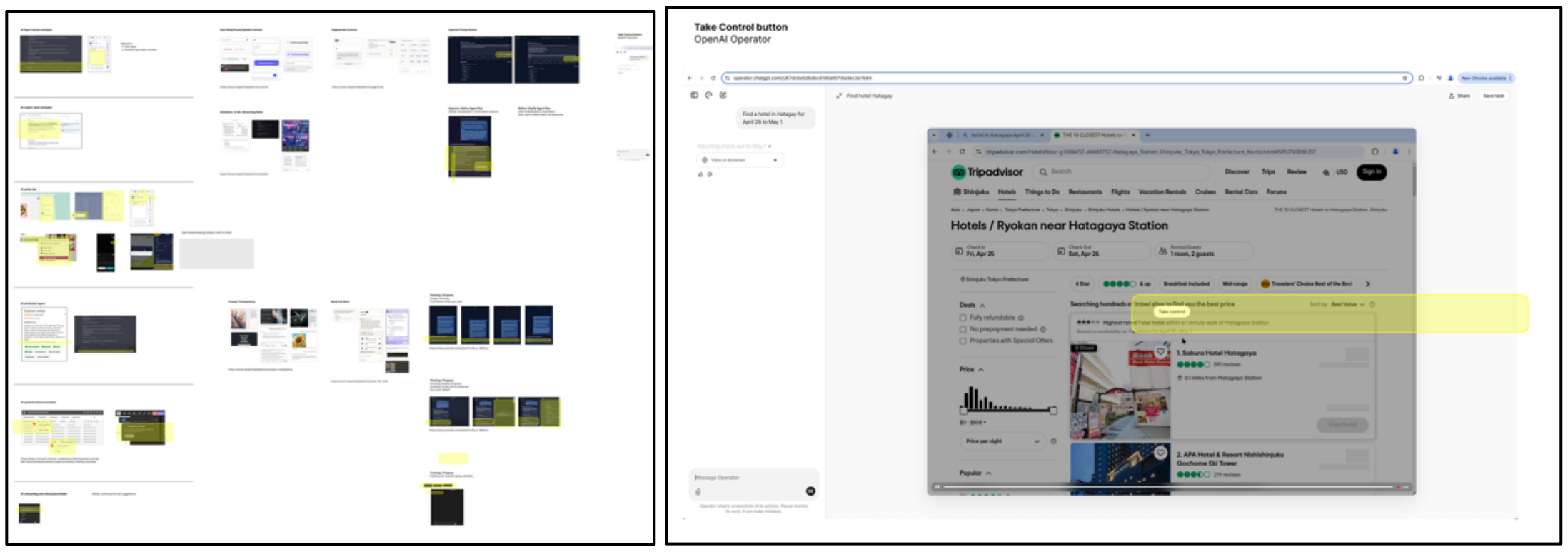

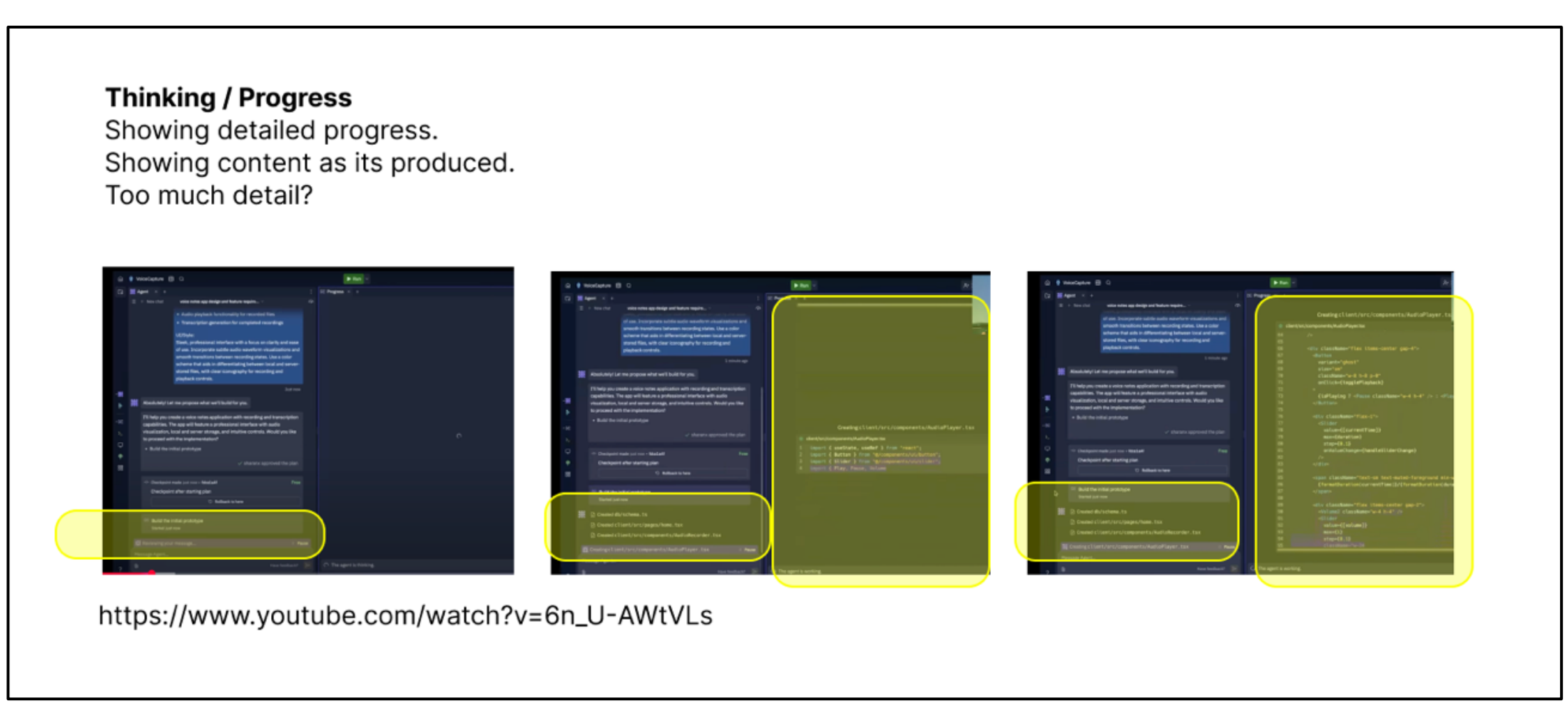

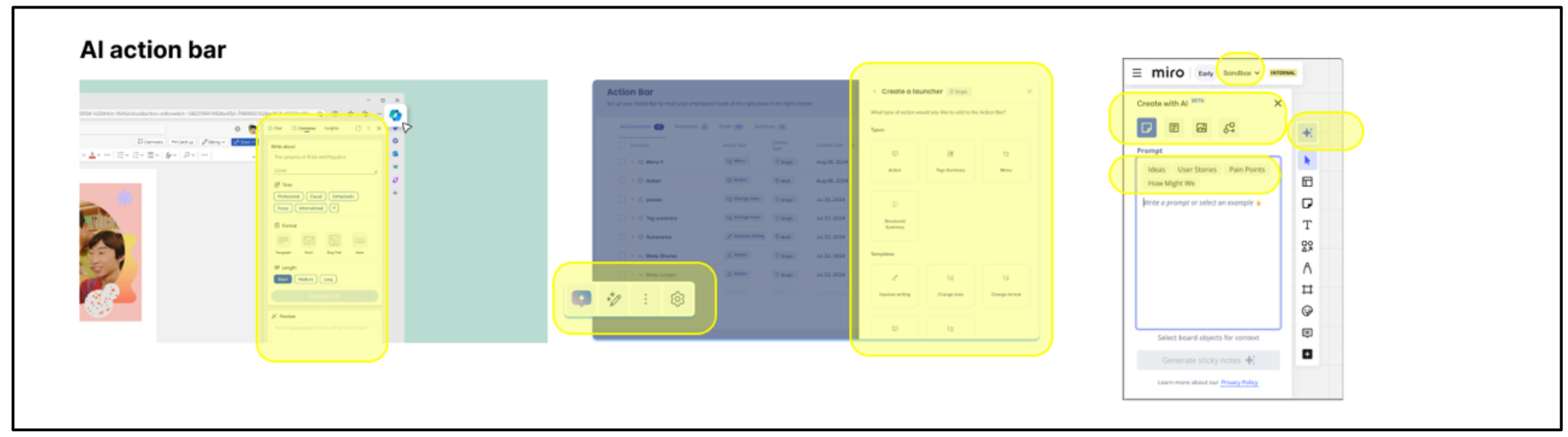

Examples of our analysis of AI application UI, grouped into themes of AI-related components, patterns, and features.

### A.1.3 AI UX Knowledge Base

We built an internal AI UX design knowledge base for AWS. We compiled terminology, principles, guidelines, and design patterns from varied sources, including online pattern libraries, articles for UX practitioners, and research publication. Sample pages are provided below.

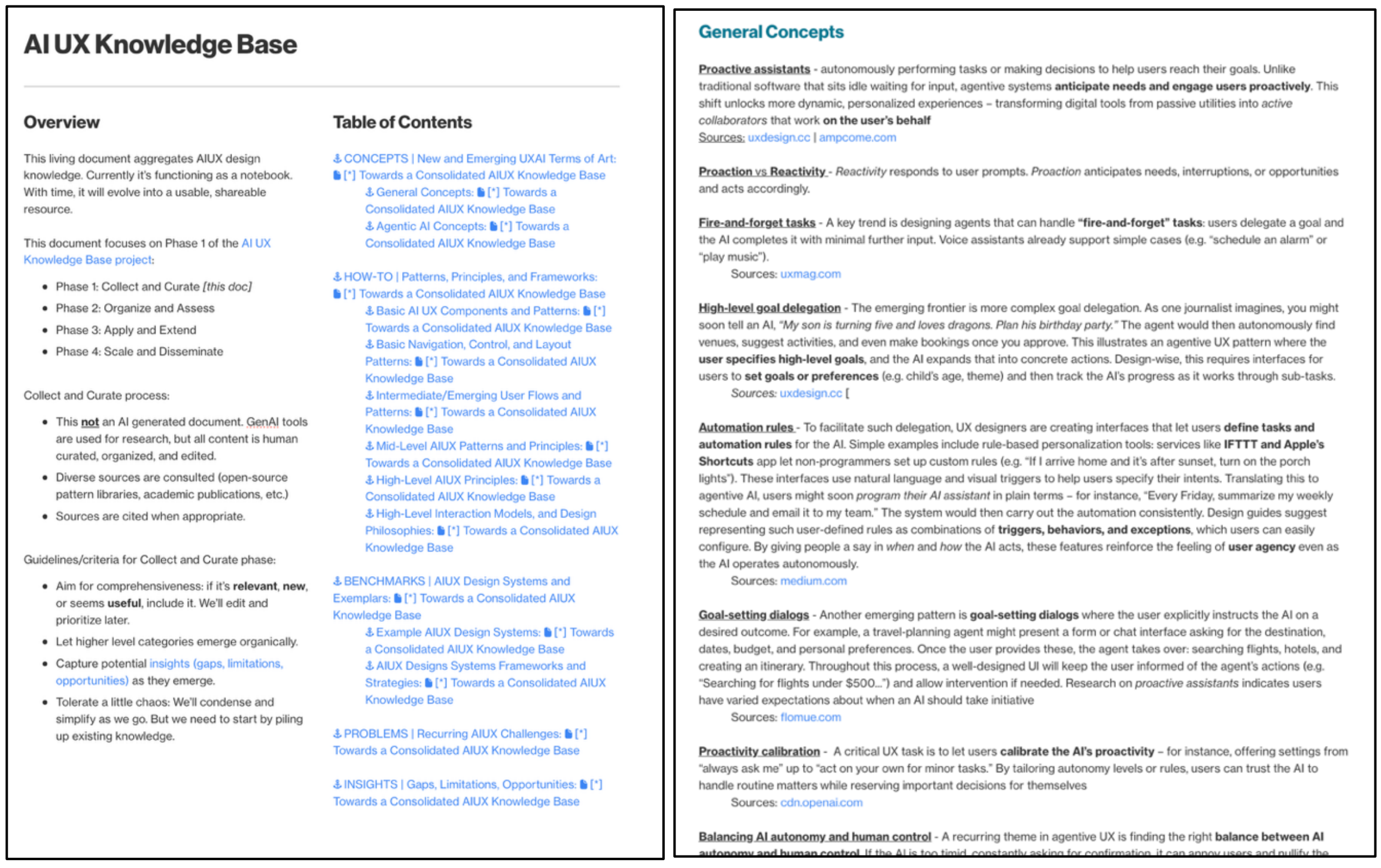

# AI UX Knowledge Base

## Overview

This living document aggregates AIUX design knowledge. Currently it's functioning as a notebook. With time, it will evolve into a usable, shareable resource.

This document focuses on Phase 1 of the AI UX Knowledge Base project:

- Phase 1: Collect and Curate *[this doc]*
- Phase 2: Organize and Assess
- Phase 3: Apply and Extend
- Phase 4: Scale and Disseminate

Collect and Curate process:

- This **not** an AI generated document. GenAI tools are used for research, but all content is human curated, organized, and edited.
- Diverse sources are consulted (open-source pattern libraries, academic publications, etc.)
- Sources are cited when appropriate.

Guidelines/criteria for Collect and Curate phase:

- Aim for comprehensiveness: if it's **relevant**, **new**, or seems **useful**, include it. We'll edit and prioritize later.
- Let higher level categories emerge organically.
- Capture potential insights (gaps, limitations, opportunities) as they emerge.
- Tolerate a little chaos: We'll condense and simplify as we go. But we need to start by piling up existing knowledge.

## Table of Contents

## General Concepts

**Proactive assistants** - autonomously performing tasks or making decisions to help users reach their goals. Unlike traditional software that sits idle waiting for input, agentive systems **anticipate needs and engage users proactively**. This shift unlocks more dynamic, personalized experiences – transforming digital tools from passive utilities into *active collaborators* that work **on the user's behalf**
Sources: uxdesign.cc | ampcome.com

**Proaction vs Reactivity** - *Reactivity* responds to user prompts. *Proaction* anticipates needs, interruptions, or opportunities and acts accordingly.

**Fire-and-forget tasks** - A key trend is designing agents that can handle **"fire-and-forget" tasks**: users delegate a goal and the AI completes it with minimal further input. Voice assistants already support simple cases (e.g. "schedule an alarm" or "play music").
Sources: uxmag.com

**High-level goal delegation** - The emerging frontier is more complex goal delegation. As one journalist imagines, you might soon tell an AI, *"My son is turning five and loves dragons. Plan his birthday party."* The agent would then autonomously find venues, suggest activities, and even make bookings once you approve. This illustrates an agentive UX pattern where the **user specifies high-level goals**, and the AI expands that into concrete actions. Design-wise, this requires interfaces for users to **set goals or preferences** (e.g. child's age, theme) and then track the AI's progress as it works through sub-tasks.
*Sources:* uxdesign.cc [

**Automation rules** - To facilitate such delegation, UX designers are creating interfaces that let users **define tasks and automation rules** for the AI. Simple examples include rule-based personalization tools: services like **IFTTT and Apple's Shortcuts** app let non-programmers set up custom rules (e.g. "If I arrive home and it's after sunset, turn on the porch lights"). These interfaces use natural language and visual triggers to help users specify their intents. Translating this to agentive AI, users might soon *program their AI assistant* in plain terms – for instance, "Every Friday, summarize my weekly schedule and email it to my team." The system would then carry out the automation consistently. Design guides suggest representing such user-defined rules as combinations of **triggers, behaviors, and exceptions**, which users can easily configure. By giving people a say in *when* and *how* the AI acts, these features reinforce the feeling of **user agency** even as the AI operates autonomously.
Sources: medium.com

**Goal-setting dialogs** - Another emerging pattern is **goal-setting dialogs** where the user explicitly instructs the AI on a desired outcome. For example, a travel-planning agent might present a form or chat interface asking for the destination, dates, budget, and personal preferences. Once the user provides these, the agent takes over: searching flights, hotels, and creating an itinerary. Throughout this process, a well-designed UI will keep the user informed of the agent's actions (e.g. "Searching for flights under $500...") and allow intervention if needed. Research on *proactive assistants* indicates users have varied expectations about when an AI should take initiative
Sources: flomue.com

**Proactivity calibration** - A critical UX task is to let users **calibrate the AI's proactivity** – for instance, offering settings from "always ask me" up to "act on your own for minor tasks." By tailoring autonomy levels or rules, users can trust the AI to handle routine matters while reserving important decisions for themselves
Sources: cdn.openai.com

**Balancing AI autonomy and human control** - A recurring theme in agentive UX is finding the right **balance between AI autonomy and human control**. If the AI is too timid, constantly asking for confirmation, it can annoy users and nullify the

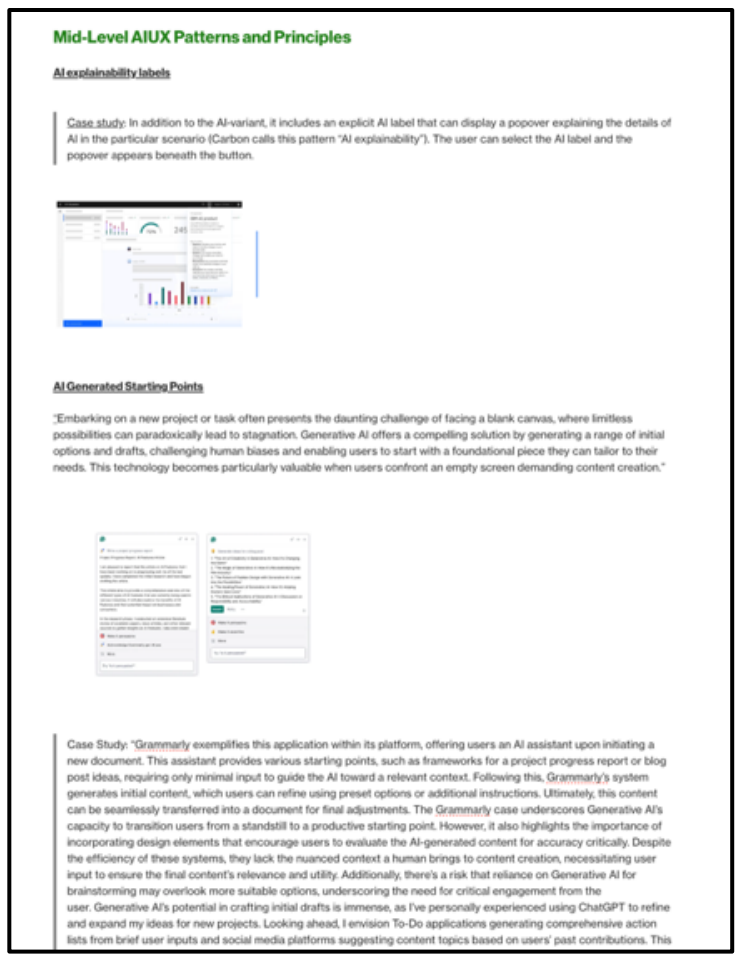

# *HOW-TO* | Patterns, Principles, and Frameworks

This sections collects design patterns, principles, and frameworks for applying actionable design knowledge to recurring AIUX problems and opportunities.

## Basic AI UX Components and Patterns

**AI input canvas** - The input fields for prompts (text, voice, image, etc.)

**AI output cards** - Modular containers for AI-generated content.

**AI output fields** - More generally, any AI output, which may not be constrained to a card. Example: sends a mes triggers a menu opening, updates UX copy elsewhere

**AI action bar** - A toolbar for pre- or post-generation actions (refine, share, export, etc.). They often appear as a contextual menu with actions such as "Summarize," "Rewrite"," "Translate," and "Generate Images."

**AI attribution layer** - Display of sources, limitations, model provenance, confidence, etc.

### *A.1.4 AI UX Knowledge Base Content*

13 AI UX resources we reviewed in detail.

| Site Name and URL (last accessed September 2025) | Website Screenshot |
|---|---|
| Shape of AI | https://www.shapeof.ai/ |
| AIVerse | https://aiverse.design/ |
| Smarter Patterns | https://smarterpatterns.com/patterns |
| HAX Toolkit (Microsoft) | https://www.microsoft.com/en-us/haxtoolkit/library/ |
| People + AI Guidebook Patterns (Google) | https://pair.withgoogle.com/guidebook/patterns |
| Maggie Appleton Design Patterns | https://maggieappleton.com/patterns |
| AI UX Patterns | https://www.aiuxpatterns.com/ |
| UX for UI | https://www.uxforai.com/archive?tags=Design+Patterns |
| IBM Design for AI | https://www.ibm.com/design/ai/ |
| AI Product UX Patterns | https://aiux.rezza.io/ |
| IF Design Patterns | https://catalogue.projectsbyif.com/ |
| Data Cards | https://sites.research.google/datacardsplaybook/patterns/ |
| Google AI Principles | https://ai.google/responsibility/principles/ |